\documentclass[11pt]{article}

\usepackage{amssymb}
\usepackage{amsmath}
\usepackage{paralist}
\usepackage{graphicx}
\usepackage{wrapfig}
\usepackage{xspace}
 \usepackage{url}
 \usepackage{color}
 \definecolor{blu}{rgb}{.3,0,.3}
\usepackage[colorlinks,linkcolor=blue,filecolor=blue,citecolor=blue,urlcolor=blue,pdfstartview=FitH,plainpages=true]{hyperref}
\hypersetup{colorlinks,linkcolor=[rgb]{.3,0,.3},filecolor=blue,citecolor=blue,urlcolor=green,pdfstartview=FitH,plainpages=true}
\usepackage{tikz}
\usetikzlibrary{graphs,graphs.standard,shapes,calc}
\usepackage{soul}
\usepackage{algorithm}
\usepackage{algorithmic}

\newtheorem{theorem}{Theorem}[section]

\newtheorem{definition}{Definition}[section]

\newtheorem{lemma}[theorem]{Lemma}

\newtheorem{corollary}[theorem]{Corollary}

\newcommand{\paper}{\text{paper}\xspace}

\newcommand{\dsection}{\text{section}\xspace}

\newcommand{\dsubsection}{\text{section}\xspace}

\newcommand{\sectionref}[1]{\namedref{Sec.}{#1}}
\newcommand{\subsectionref}[1]{\namedref{Sec.}{#1}}

\newcommand{\boaz}[1]{\hl{\textbf{BPS}: #1}}

\newcommand{\todo}[1]{{\color{violet}{\bfseries\{TODO: #1\}}}}
\newcommand\ignore[1]{}

\newcommand{\namedref}[2]{{#1~\ref{#2}}}

\newcommand{\theoremref}[1]{\namedref{Theorem}{#1}}
\newcommand{\defref}[1]{\namedref{Definition}{#1}}
\newcommand{\figref}[1]{\namedref{Fig.}{#1}}
\newcommand{\lemmaref}[1]{\namedref{Lemma}{#1}}

\newcommand{\corollaryref}[1]{\namedref{Cor.}{#1}}

\newcommand{\algref}[1]{\namedref{Alg.}{#1}}
\newcommand{\equalityref}[1]{\namedref{Eq.}{#1}}

\newcommand{\stepref}[1]{\hyperref[#1]{Step~(\ref*{#1})}}

\def\blackslug{\hbox{\hskip 1pt \vrule width 4pt height 8pt
		depth 1.5pt \hskip 1pt}}
\def\QED{\quad\blackslug\lower 8.5pt\null\par}

\def\inQED{~~~~~\quad\blackslug\lower 8.5pt\null}
\def\Proof{\noindent{\bf Proof:~}}

\newcommand{\Unit}{\tilde{\mathbf{1}}}

\newcommand{\Clus}{\mathrm{cl}}
\newcommand{\Val}{\mathrm{vl}}

\newcommand{\CC}{\mathcal{C}}

\newcommand{\Set}[1]{\left\{ #1 \right\}}

\newcommand{\ceil}[1]{\left\lceil #1 \right\rceil}
\newcommand{\floor}[1]{\left\lfloor #1 \right\rfloor}

\def\DEF{\stackrel{\rm def}{=}}
\newcommand{\Eqr}[1]{\equalityref{#1}}  

\newcommand\cwc{\textrm{CWC}\xspace}
\def\CWC{\cwc}
\newcommand\argmin{\operatorname*{argmin}}
\newcommand\argmax{\operatorname*{argmax}}
\newcommand\WLOG{{w.l.o.g.}\xspace}
\newcommand\bl{b_\ell}
\newcommand\bc{b_c}
\newcommand\blmin{\phi}
\newcommand\bC[1]{\bc({#1})}
\newcommand\bL[2]{w({{#1}, {#2}})}
\newcommand\kkR[2]{k_{#2}({#1})}
\newcommand\kR[1]{k({#1})}
\newcommand\IR[1]{I_{#1}}
\newcommand\ZR[1]{Z_{#1}}

\newcommand\interval[2]{[{#1}, {#2}]}
\newcommand\kCR[1]{k_c(#1)}
\newcommand\kLR[1]{k_\ell(#1)}
\newcommand\bLminR[1]{\blmin\left(\IR{#1} \right)}

\newcommand\Br[2]{B_{#1}({#2})}
\newcommand\BR[1]{B_{#1}}
\newcommand\Diam[0]{\mathrm{diam}}
\newcommand\load[0]{\mathrm{load}}
\newcommand\Zmax[0]{{Z_{\max}}}
\newcommand\ZC[0]{{Z}}

\newcommand{\CCW}{\textsf{cComb}\xspace}
\newcommand{\CR}{\textsf{cR}\xspace}
\newcommand{\CW}{\textsf{cW}\xspace}
\newcommand{\CAW}{\textsf{cAW}\xspace}
\newcommand{\CAR}{\textsf{cAR}\xspace}
\newcommand{\CCR}{\textsf{cCast}\xspace}
\newcommand{\FW}{\texttt{FW}\xspace}
\newcommand{\FR}{\texttt{FR}\xspace}

\newcommand{\para}[1]{\par\noindent$\blacktriangleright$~\emph{#1}}
\renewcommand{\paragraph}[1]{\par\medskip\noindent\textbf{#1}}
\def\tp{\!+\!}
\def\tm{\!-\!}
\def\cC{\mathcal{C}}

\usepackage{graphicx,accents}

\makeatletter
\DeclareRobustCommand{\cev}[1]{%
	\mathpalette\do@cev{#1}%
}
\newcommand{\do@cev}[2]{%
	\fix@cev{#1}{+}%
	\reflectbox{$\m@th#1\vec{\reflectbox{$\fix@cev{#1}{-}\m@th#1#2\fix@cev{#1}{+}$}}$}%
	\fix@cev{#1}{-}%
}
\newcommand{\fix@cev}[2]{%
	\ifx#1\displaystyle
	\mkern#23mu
	\else
	\ifx#1\textstyle
	\mkern#23mu
	\else
	\ifx#1\scriptstyle
	\mkern#22mu
	\else
	\mkern#22mu
	\fi
	\fi
	\fi
}

\newcommand\FG[2]{FG_{#2}({#1})}

\newcommand\QstF[0]{Quickest Flow\xspace}
\newcommand{\val}{\mathrm{value}}

\title{\textbf{Distributed Computing With the Cloud}}

\author{\begin{tabular}{ccc}
   Yehuda Afek & Gal Giladi & Boaz Patt-Shamir \\
		\multicolumn{2}{c}{School of CS}&{School of EE}\\
		\multicolumn{3}{c}{Tel Aviv University}\\ 
		\multicolumn{3}{c}{Tel Aviv 6997801}\\ 
		\multicolumn{3}{c}{Israel}
	\end{tabular}
}

\begin{document}
\maketitle

\begin{abstract}


We investigate the effect of omnipresent cloud storage on distributed
computing.  We specify a network model with links of prescribed
bandwidth that connect standard processing nodes, and, in addition,
passive storage nodes. Each passive node represents a cloud storage
system, such as Dropbox, Google Drive etc.  We study a few tasks in
this model, assuming a single cloud node connected to all other nodes,
which are connected to each other arbitrarily.  We give
implementations for basic tasks of collaboratively writing to and
reading from the cloud, and 
for more advanced applications such as matrix multiplication and
federated learning.  Our results show that utilizing
node-cloud links as well as node-node links can considerably speed up
computations, compared to the case where processors communicate
either only through the cloud or only through the network links.

We provide results for general directed graphs, and for graphs with
``fat'' links between processing nodes. For the general case, we
provide optimal algorithms for uploading and downloading files using
flow techniques. We use these primitives to derive algorithms for
\emph{combining}, where every processor node has an input value and
the task is to compute a combined value under some given associative
operator.  In the case of fat links, we assume that links between
processors are bidirectional and have high bandwidth, and we give
near-optimal algorithms for any commutative combining operator (such
as vector addition).  For the task of matrix multiplication (or other
non-commutative combining operators), where the inputs are ordered, we
present sharp results in the simple ``wheel'' network, where procesing
nodes are
arranged in a ring, and are all
connected to a single cloud node. 
\end{abstract}


\section{Introduction}
\label{sec:intro}

\ignore{
\paragraph{Background and the new model.}
}

In 2018 Google announced that the number of users of
Google Drive is surpassing one billion~\cite{google-drive}. 
Earlier that year, Dropbox stated that in total, more than an exabyte
($10^{18}$ bytes) of data has been uploaded by its 
users~\cite{dropbox}.
Other cloud-storage services, such as Microsoft's OneDrive, Amazon's
S3, or Box, are thriving too.
The driving force of this paper is our
wish to let \emph{other} distributed systems to take advantage
of the enormous infrastructure that makes up the complexes called
``clouds.''  Let us explain how.

The computational and storage capacities of servers in cloud services
are relatively well advertised.  A lesser known fact is
that 
a cloud system also entails a massive component of
\emph{communication}, that makes it appear close almost everywhere on
the Internet. (This feature is particularly essential for cloud-based
video conferencing applications, such as Zoom, Cisco's Webex and
others.)  In view of the existing cloud services, our fundamental idea
is to \emph{abstract a complete cloud system as a single, passive storage
node}.

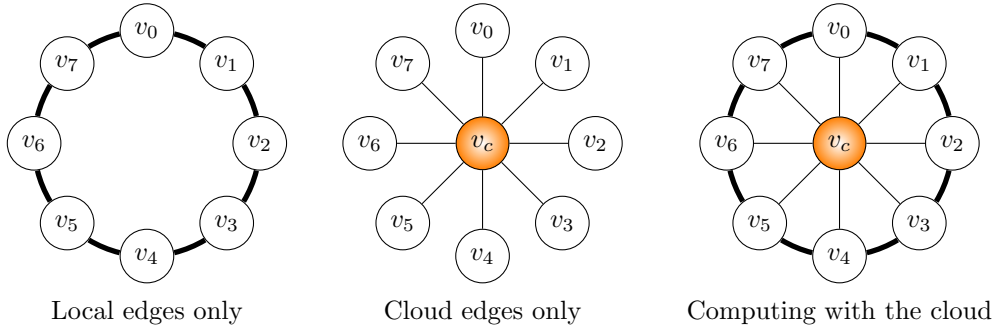
\begin{figure*}[t]
  \small
  \centering
  \def \n {7}
  \def \np {(\n+1)}
  \def \radius {1.5cm}
  \def \margin {14}
  \begin{tikzpicture}
    \node (mid) {};
    \foreach \s in {0,...,\n} {
      \node[draw, circle] (v\s) at ({90-360/\np * (\s )}:\radius) {$v_\s$};
      \draw[line width=2pt] ({360/(\np) * (\s )+\margin}:\radius) 
      arc ({360/\np * (\s)+\margin}:{360/\np * (\s+1)-\margin}:\radius);
    }%
    \node at (-90:1.5*\radius) (bot) {Local edges only};
  \end{tikzpicture}
  \hspace*{5mm}
  \begin{tikzpicture}
    \node[draw, circle, inner color=white,outer color=orange] (mid) {$v_c$};
    \foreach \s in {0,...,\n} {
      \node[draw, circle] (v\s) at ({90-360/\np * (\s )}:\radius) {$v_\s$};
      \draw[-] (mid) -- (v\s);
    }%
    \node at (-90:1.5*\radius) (bot) {Cloud edges only};
  \end{tikzpicture}
  \hspace*{5mm}
  \begin{tikzpicture}
    \node[draw, circle, inner color=white,outer color=orange] (mid) {$v_c$};
    \foreach \s in {0,...,\n} {
      \node[draw, circle] (v\s) at ({90-360/\np * (\s )}:\radius) {$v_\s$};
      \draw[line width=2pt] ({360/(\np) * (\s )+\margin}:\radius) 
      arc ({360/\np * (\s)+\margin}:{360/\np * (\s+1)-\margin}:\radius);
      \draw[-] (mid) -- (v\s);
    }%
    \node at (-90:1.5*\radius) (bot) {Computing with the cloud};
  \end{tikzpicture}
  \caption{\it Wheel topology with $n=8$.  The $v_i$ nodes are processing
    nodes connected by a ring of high-bandwidth links. The cloud node
    $v_c$ is connected to the processing nodes by lower-bandwidth
    links. All links are bidirectional and symmetric.
  }
  \label{fig-ring}
\end{figure*}

To see the benefit of this appraoch, consider a network of the ``wheel'' topology:
a single cloud node is connected to $n$ processing nodes arranged
in a cycle (see \figref{fig-ring}).
Suppose each processing node has a wide link of bandwidth $n$ to its
cycle neighbors, 
and a narrower link of bandwidth $\sqrt{n}$ to the cloud node.
Further suppose that each processing node has an $n$-bit vector,
and that the goal is to calculate the sum of all vectors.
Without the cloud (\figref{fig-ring}, left),
such a task requires at least $\Omega(n)$ rounds --
to cover the distance;
on the other hand, without using the
cycle links (\figref{fig-ring}, middle),
transmitting a single vector from any processing node (and hence computing
the sum) requires
$\Omega(n/\sqrt{n}) = \Omega(\sqrt{n})$ rounds -- due to the limited bandwidth to the cloud.
But using both cloud links and local links (\figref{fig-ring}, right),
the sum can be computed in $\tilde{\Theta} (\sqrt[4]{n} )$ rounds, as
we show in this \paper.

More generally, in this paper we initiate the study of the question of how to
use an omnipresent cloud storage to speed up computations, if possible.
We stress that the idea here is to develop a framework and tools
that facilitate 
computing \emph{with} the cloud, as opposed to computing
\emph{in} the cloud.

Specifically, in this paper we introduce the \emph{computing with the
  cloud} model (\CWC), and present algorithms that
efficiently combine distributed inputs to compute various functions,
such as vector addition and matrix multiplication. To this end, we
first implement  (using dynamic flow techniques) primitive operations
that allow for the exchange of large messages between processing nodes
and cloud nodes.
Given the combining algorithms, we show how to implement some
applications such as federated learning and file 
de-duplication (dedup).

\subsection{Model Specification}
The ``Computing with the Cloud'' (\CWC) model is a synchronous network
whose underlying topology is described by a weighted directed graph
$G=(V, E, w)$. The node set consists of two disjoint subsets:
$V=V_p\cup V_c$, where $V_p$ is the set of \emph{processing nodes},
and $V_c$ is the set of \emph{cloud nodes}.  Cloud nodes are passive
nodes that function as shared storage: they support read and write
requests, and do not perform any other computation.  
We use $n$ to denote the number
of processing nodes (the number of cloud nodes is typically constant).
	

We denote the set of links that connect two processing nodes by $E_L$
(``local links''), and by $E_C$ (``cloud links'') the set of links
that connect processing nodes to cloud nodes.  Each link
$e\in E=E_L\cup E_C$ has a prescribed bandwidth $w(e)$ (there are no
links between different cloud nodes).  We denote by $G_p\DEF(V_p,E_L)$
the graph $G-V_c$, i.e., the graph spanned by the processing nodes.

Our execution model is the standard synchronous network model,
where each round consists of processing nodes receiving messages sent
in the previous round, doing an arbitrary local computation, and then
sending messages.  The size of a message sent over a link $e$ in a
round is at most $w(e)$ bits.
 

Cloud nodes do not perform any computations: they can only receive
requests we denote by \FR and \FW (file read and write, respectively)
, to which they respond in the following round. More precisely, each
cloud node has unbounded storage; to write, a processing node $v_i$
invokes $\FW$ with arguments that describe the target cloud node, a
filename $f$, a bit string $S$, and the location (index) within $f$
that $S$ needs to be written in. It is assumed that
$|S|\le w(v_i,v_c)$ bits (longer writes can be broken to a series of
\FW operations). To read, a processing node $v_i$ invokes $\FR$ with
arguments that describe the cloud node, a filename $f$ and the range
of indices to fetch from $f$. Again, we assume that the size of the
range in any single \FR invocation by node $v_i$ is at most
$w(v_i,v_c)$.%
\footnote{ For both the \FW and \FR operations we ignore the metadata
  (i.e., $v_c$'s descriptor, the filename $f$ and the indices) and
  assume that the total size of metadata in a single round is
  negligible and can fit within $w(v_i,v_c)$.  Otherwise, processing
  nodes may use the metadata parameters to exchange information that
  exceeds the bandwidth limitations (for example, naming a file with
  the string representation of a message whose length is larger than
  the bandwidth).  }

\FW operations are exclusive, i.e., no other operation (read or
write) to the same file location is
allowed to take place simultaneously.
Concurrent \FR operations from the same location are allowed.

\smallskip
\noindent\textbf{Discussion.}
We believe that our model is fairly widely applicable.
A processing node in our model may represent anything from
a computer cluster with a single gateway to the Internet, to
cellphones or even smaller devices---anything with a non-shared Internet
connection. The local links can range from high-speed fiber to
Bluetooth or infrared links. Typically in this setting the local links
have bandwidth much larger than the cloud links (and cloud downlinks
in many cases have larger bandwidth than cloud uplinks).
Another possible interpretation of the model
is a private network (say, in a corporation), 
where a cloud node represents a storage or a file
server. In this case the cloud link bandwidth may be as large as the local
link bandwidth.

\subsection{Problems Considered and Main Results}
\label{sec:problem-statements}
Our main results in this \paper 
are efficient algorithms in
the \CWC model to combine values stored at nodes
.  These algorithms use building blocks that facilitate efficient
transmission of large messages between  processing nodes and
cloud nodes. These building blocks, in turn, are implemented
in a straightforward way using dynamic flow techniques.  Finally, we
show how to use the combining algorithms to derive new algorithms for
federated learning and file de-duplication (dedup) in the \CWC model.
More specifically, we provide implementations of the following tasks.

\noindent\emph{Basic cloud operations:}
Let  $v_c$ denote a cloud node below.
\begin{compactitem}
\item 
\emph{$\CW_i$ (cloud write):} write an $s$-bits file $f$ stored at
node $i\in V_p$ to  node $v_c$.
\item 
\emph{$\CR_i$ (cloud read):} fetch an $s$-bits file $f$
from node $v_c$ to node $i\in V_p$. 
\item 
\emph{$\CAW$ (cloud all write):} for each $i\in V_p$, write an $s$-bits file $f_i$ stored at
node $i$ to  node $v_c$.
\item 
\emph{$\CAR$ (cloud all read):} for each $i\in V_p$, fetch an $s$-bits file $f_i$
from node $v_c$ to node $i$. 
\end{compactitem}

\noindent\emph{Combining and dissemination operations:}
\begin{compactitem} 
\item 
\emph{$\CCW$: (cloud combine):} Each node $i\in V_p$ has an $s$-bits input string $S_i$, and
there is a binary associative operator
$\otimes:\Set{0,1}^s\times\Set{0,1}^s\to\Set{0,1}^s$
(the result is as long as each  operand). The requirement
is to write to a cloud node $v_c$ the $s$-bits string 
$S_1\otimes S_2\otimes\cdots\otimes S_n$.
 Borrowing from Group Theory, we call the operation
$\otimes$ \emph{multiplication}, and $S_1\otimes S_2$ is
the \emph{product} of $S_1$ by $S_2$. In general, $\otimes$ is not
necessarily commutative. 
We assume the existence of a
unit element for $\otimes$, denoted $\Unit$, such that
$\Unit\otimes S=S\otimes\Unit=S$ for any $s$-bits strings $S$. The
unit element is represented by a string of $O(1)$ bits.
Examples for commutative operators include vector (or matrix) addition
over a finite field, logical bitwise
operations, 
leader election, 
and the top-$k$ problem. 
Examples for non-commutative operators may be matrix multiplication
(over a finite field) and function
composition. 

\item 
\emph{\CCR (cloudcast):} All the nodes $i\in V_p$ simultaneously fetch a copy of an $s$-bits file $f$
from  node $v_c$. (Similar to network broadcast.)
\end{compactitem}

\paragraph{Applications.}
 \CCW and \CCR can be used directly to provide
matrix multiplication, matrix addition, and vector addition. We also
outline the implementation of the following.

\noindent
\emph{Federated learning (FL) \cite{federated-17}:} In FL, a
collection of agents collaborate in training a 
neural network to construct a model of some concept, but the agents
 want to keep their data private.
Unlike \cite{federated-17}, in our model the central server is 
a passive storage device that does not carry out computations.
We show how  elementary secure computation techniques, along with our
combining algorithm, can efficiently help  training an ML model in the federated scheme
implemented in \CWC, while maintaining privacy.

\noindent
\emph{File deduplication:} Deduplication (or dedup) is a task in file
stores, where redundant identical copies of data are identified (and
possibly unified)---see, e.g.,~\cite{dedup-performance}.  Using \CCW
and \CCR, we implement file dedup in the \CWC model on collections of
files stored at the different processing nodes. The algorithm keeps
a single copy of each file and pointers instead of the other replicas.

\paragraph{Special topologies.}
The complexity of the general algorithms we present depends on the
given network topology. We study a few cases of interest.  

First, we consider \emph{$s$-fat-links} network, defined to be, for
a given parameter $s\in\mathbb{N}$, as the \CWC model with the following additional assumptions:
\begin{compactitem}
\item All links are symmetric, i.e., $w(u,v)=w(v,u)$ for every link $(u,v) \in E$.
\item Local links have bandwidth at least $s$.
\item There is only one cloud node $v_c$. 
\end{compactitem}
The fat links model seems suitable in many real-life cases where
local links are much wider than cloud links (uplinks to the Internet), as is the intuition
behind the HYBRID model \cite{AHKSS20}.

Another topology we consider is the \emph{wheel network}, depicted
schematically in \figref{fig-ring} (right).
In a wheel system there are $n$
processing nodes 
arranged in a ring, and a cloud
node 
connected to  all processing nodes.
%
In the \emph{uniform} wheel, all cloud links have the same bandwidth
$b_c$ and all local links have the same bandwidth $b_l$. In the
uniform wheel model, we typically assume that $\bc\ll\bl$.

The wheel network is motivated by non-commutative combining
operations, where the order of the operands induces a linear order on
the processing nodes, i.e., we view the nodes as a line, where the
first node holds the first input, the second node holds the second
input etc. For symmetry, we connect the first and the last node, and
with a cloud node connected to all---we've obtained the wheel.

\paragraph{Overview of techniques.}
As mentioned above, the basic file operations (\CW, \CR, \CAW and
\CAR) are solved optimally using {dynamic flow} techniques, or more
specifically, \emph{quickest flow} (\sectionref{sec:rw}%
, which have been studied in
 numerous papers in the past (cf.~\cite{quickest,Skutella-flows}).
We present closed-form bounds on \CW and \CR for the wheel topology
in \sectionref{sec:wheel}.

We present tight 
bounds for \CW and \CR in the {$s$-fat-links} network, where $s$ is
the input size at all nodes.  We then continue to consider the tasks
\CCW with \emph{commutative operators} and \CCR, and
prove nearly-tight bounds on their time complexity in the
{$s$-fat-links} network (\theoremref{thm-ccw-ub},
\theoremref{theorem:CCW-LB}, \theoremref{thm-AR-LB}).
The idea is to first find, for every processing node $i$, a cluster
of processing nodes that allows it to perform \CW in an optimal number
of rounds.  We then perform \CCW by combining the values within every
cluster using convergecast \cite{Peleg:book}, and then combining the
results in a computation-tree fashion.  Using sparse
covers~\cite{AP90}, we  perform the described procedure in near-optimal
time. 

\emph{Non-commutative operators} are explored in the natural wheel
topology. We present algorithms for wheel networks with
\emph{arbitrary} bandwidth (both cloud and local links).
We prove an upper bound 
for \CCW (\theoremref{thm-ccw-intervals-ub} ) and a lower bound of
(\theoremref{thm-aw-lb}).

Finally, in \sectionref{sec:applications}, we demonstrate how the considered tasks
can be applied for the purposes of Federated Learning and File Deduplication.

\paragraph{Paper organization.} 
In \sectionref{sec:rw} we study the topology of basic primitives  in the
\CWC model.  %
In \sectionref{sec:va} we study combining algorithms in
general topologies in fat links networks. %
In \sectionref{sec:wheel} we consider combining for non-commutative
operators in the wheel topology. %
In \sectionref{sec:applications} we discuss application level usage
of the \CWC model, such as federated learning and
deduplication. Conclusions and open problems are presented in
\sectionref{sec:conc}.

\subsection{Related Work}
\label{app-related}
Our model is based on, and inspired by, a long history of theoretical
models in distributed computing. To gain some perspective, we offer
here a brief review.

Historically, distributed computing is split along the dichotomy of
message passing vs shared memory~\cite{Pierre10}. While message
passing is deemed the ``right'' model for network algorithms, the
shared memory model is the abstraction of choice for programming
multi-core machines. 

The prominent message-passing models are LOCAL~\cite{L92},
and its derived CONGEST~\cite{Peleg:book}.
(Some models also include a broadcast channel, e.g.~\cite{ALSY}.) In
both LOCAL and CONGEST, a system
is represented by a connected (typically undirected) graph, in which
nodes represent processors and edges represent communication links.
\ignore{
The basic actions are message send (initiated by processors) and
message receive (initiated by links). Local computation is assumed to
be unbounded and instantaneous, so that the running time of a network
algorithm is just the number of communication rounds it requires.
In the LOCAL model, message size is unbounded, and therefore the only
restriction for a node to retrieve any piece of information  is
the number of hops separating the node from the location of that
information.
%
The CONGEST model is a refined version of LOCAL, with the additional
restriction that messages may be only $O(\log n)$-bits long, where $n$ is the number of
nodes.%
\footnote{%
	More precisely, the number of bits in a message is a parameter of
	the CONGEST model, and the default value of that parameter is
	$O(\log n)$.  This value is chosen for reasons similar to the choice
	of word size in the sequential RAM model~\cite{AhoHU74}: this way, a
	message can contain a constant number of references to nodes and
	edges, and possibly other
	variables whose magnitude is bounded by a polynomial in $n$.
}
} In LOCAL, message  size is unbounded, while in CONGEST, message
size is restricted, typically to $O(\log n)$ bits.
Thus, CONGEST 
accounts not only for the
distance information has to traverse, but also for information volume  
and  the
bandwidth available for its transportation.

While most algorithms in the LOCAL and CONGEST models assume
fault-free (and hence synchronous) executions, in the distributed
shared memory model, asynchrony and faults are the primary source of
difficulty. Usually, in the shared memory model one assumes that there
is a collection	 of ``registers,'' accessible by multiple threads of
computation that run at different speeds and may suffer crash or even
Byzantine faults (see, e.g., \cite{AttiyaW}). The main issues in this
model are coordination and fault-tolerance. Typically, the only
quantitative hint to communication cost is the number and size of the
shared registers.

Quite a few papers consider the combination of message
passing and shared memory, e.g., 
\cite{MNV,FKK,logp-new,qsm,BSP,AdlerGMR-99}.
The uniqueness of the \CWC model with respect to past work is that it  combines passive storage nodes with a  message passing network with restrictions on the links bandwidth.
The CONGESTED CLIQUE (CC) model~\cite{LotkerPPP-05}
is a special case of CONGEST,
where the underlying 
graph is assumed to be fully
connected. 
The CC model is appropriate for computing \emph{in} the cloud, as it
has been shown that under some relatively mild
conditions, algorithms designed for the CC 
model can be implemented in the MapReduce model, i.e., run in 
datacenters~\cite{HegemanP-15}. Another model for computing in the
cloud is the MPC model~\cite{KSV10}. 
%
Very recently, the HYBRID model~\cite{AHKSS20} was proposed as a 
combination of CC
with classical graph-based communication. 
More specifically, the HYBRID model assumes the existence of  two
communication networks: one for local communication between neighbors,
where links are typically of infinite bandwidth (exactly like LOCAL);
the other network is a \emph{node-congested} clique, i.e., a node can
communicate with every other node directly via ``global links,'' but
there is a small upper bound (typically $O(\log n)$) on 
the total number 
of messages a node can
send or receive via these global links in a round.
Even though the model was presented only recently, there is already 
a line of algorithmic work in it, in particular for computing
shortest paths~\cite{AHKSS20,KS-20,CLP20}.

\textbf{Discussion.} Intuitively, our  \CWC model can be viewed as
the classical 
CONGEST model over the processors, augmented by
special cloud nodes  (object stores)
connected to some (typically, many) compute nodes.
To reflect modern demands and availability of resources, we relax the 
very stringent bandwidth 
allowance of
CONGEST, and usually envision networks with much larger link
bandwidth (e.g., $n^\epsilon$ for some $\epsilon>0$).

Considering previous network models, it appears that  HYBRID  is
the closest to  \CWC, even though HYBRID
was not expressly designed to model the cloud. In our view,
\CWC is indeed more 
appropriate for computation with  the cloud. First, 
in most
cases, global communication (modeled by clique edges in HYBRID) is 
limited by link bandwidth, unlike 
HYBRID's
node capacity constraint, which seems somewhat artificial. Second, HYBRID is 
not readily amenable to  model multiple clouds, while this is a
natural property of \CWC . 

Regarding shared memory models, we are unaware of topology-based
bandwidth restriction on shared memory access in distributed
models. In some general-purpose parallel computation models (based on 
BSP~\cite{BSP}), communication capabilities are specified using a 
few
global parameters such as latency and throughput
, but these models deliberately abstract topology away.
In distributed (asynchronous) shared memory, the number of bits 
that need to be
transferred to and from the shared memory is seldom explicitly
analyzed. 



\section{Implementation of Basic Communication Primitives in \CWC}
\label{sec:rw}
\label{app:multiple}
In this section we give tight complexity results for the basic
operations of reading or writing to the cloud, by one or all
processors.  The results are derived using standard dynamic flow
techniques. We first review dynamic flows in
\sectionref{ssec:dynamic}, and then apply them to the \CWC model in
\sectionref{ssec:dyn-cwc}.

\subsection{Dynamic Flows}
\label{ssec:dynamic}
The concept of \emph{quickest flow}~\cite{quickest}, a variant of
\emph{dynamic flow}~\cite{Skutella-flows}, is defined as follows.%
\footnote{ We simplify the original definition to our context by
  setting all transmission times  to $1$.  }
A \emph{flow network}
consists of a directed weighted graph $G=(V,E,c)$ where
$c:E\to\mathbb{N}$, with a distinguished source and sink nodes,
denoted $s,t\in V$, respectively.  A \emph{dynamic flow} with
\emph{time horizon} $T\in\mathbb{N}$ and \emph{flow value} $F$ is a
mapping $f:E\times[1,T]\to \mathbb{N}$ that specifies for each edge
$e$ and time step $j$, how much flow $e$ carries between steps $j-1$
and $j$, subject to the natural constraints:
\begin{compactitem}
\item Edge capacities. For all $e\in E, j\in[1,T]$:
  \begin{equation}
    f(e,j)\le~c(e)\label{const-edge}
  \end{equation}
\item Only arriving flow can leave. For all $ v\in V\setminus\Set{s},
  j\in[1,T-1]$: 
  \begin{equation}
    \sum_{i=1}^{j}\sum_{(u,v)\in E}\!\!f((u,v),i)
  \,\ge  \sum_{i=1}^{j+1}\sum_{(v,w)\in E}\!\!f((v,w),i)
    \label{const-node}\\
  \end{equation}
\item No leftover flow. For all $v\in V\setminus\Set{s,t} $:
  \begin{equation}
    \sum_{j=1}^{T}\sum_{(u,v)\in E}\!\!f((u,v),j)
    \,=  
   \sum_{j=1}^{T}\sum_{(v,w)\in E}\!\!f((v,w),j) \label{const-node-end}
  \end{equation}
\item Flow value (source).
  \begin{equation}
  \sum_{j=1}^T\left(\sum_{(s,v)\in E}\!\!f((s,v),j)
         \!\!\,\right.\left.-\!\!
   \sum_{(u,s)\in E}\!\!f((u,s),j)\right)
   =F\label{const-source}
  \end{equation}
\item Flow value (sink).
  \begin{equation}
    \sum_{j=1}^T\left(\sum_{(t,v)\in E}\!\!f((t,v),j)
      \!\!\,\right.\left.-\!\!\sum_{(u,t)\in E}\!\!f((u,t),j)\right)
  =-F   \label{const-sink}
  \end{equation}
\end{compactitem}

In the usual flavor of dynamic
flows, $T$ is given and the goal is to maximize $F$. In the quickest
flow variant, the roles are reversed.

\begin{definition}
	Given a flow  network,
	the \textbf{quickest flow} for a given value $F$ is a dynamic flow
	$f$ satisfying (\ref{const-edge}-\ref{const-sink}) above with flow value $F$, such that the
	time horizon $T$ is minimal.
\end{definition}

\begin{theorem}[\cite{quickest}]
	\label{thm-quickest}
	The quickest flow problem can be found in strongly polynomial time.
\end{theorem}

The \emph{Evacuation problem}~\cite{evacuation-problems} is a variant
of dynamic flow that we use, specifically the case of a single
sink node \cite{Skutella-multiple-sources}.  In this problem, each
node $v$ has an initial amount of $F(v)$ flow units, and the goal is
to ship all the flow units to a single sink node $t$ in shortest
possible time (every node $v$ with $F(v)>0$ is considered a
source).  Similarly to the single source case, shipment is
described by  a
mapping $f:E\times[1,T]\to \mathbb{N}$ where $T$ is the time horizon.
The dynamic flow is subject to the edge capacity
constraints (\Eqr{const-edge}), and the following additional 
constraints:
\begin{compactitem}
\item Only initial and arriving flow can leave.  For all $ v\in V, j\in[1,T-1]$:
  \begin{equation}
    F(v)+\sum_{i=1}^{j}\sum_{(u,v)\in E}\!\!f((u,v),i)
    \,\ge
    \sum_{i=1}^{j+1}\sum_{(v,w)\in E}\!\!f((v,w),i)
	\label{const-mul-node}
  \end{equation}
\item Flow value (sources). For all $v\in V$:
  \begin{equation}
	\sum_{j=1}^T\left(\sum_{(v,w)\in E}\!\!f((v,w),j)
	-\!\!
	\sum_{(u,v)\in E}\!\!f((u,v),j)\right)
	=F(v)
	\label{const-mul-source}
  \end{equation}
\item Flow value (sink).
  \begin{equation}
	\sum_{j=1}^T\left(\sum_{(u,t)\in E}\!\!f((u,t),j)
	-\!\!
	\sum_{(t,w)\in E}\!\!f((t,w),j)
	\right)=
	\sum_{v\in V}{F(v)}
	\label{const-mul-sink}
  \end{equation}
\end{compactitem}
Formally, we use the following definition and result.

\begin{definition} \label{def-evacuation} Given a flow network in
  which each node $v$ has value $F(v)$, a solution to the evacuation
  problem is a dynamic flow of multiple sources $f$ satisfying
  (\ref{const-edge}) and (\ref{const-mul-node}-\ref{const-mul-sink}),
  such that the time horizon $T$ is minimal.
\end{definition}

\begin{theorem}[\cite{Skutella-multiple-sources}]
	\label{thm-evacuation}
	The evacuation problem can be found in strongly polynomial time.
\end{theorem}

\subsection{Using Dynamic Flows in \CWC}
\label{ssec:dyn-cwc}
In this \dsection we show how to implement  basic cloud access
primitives using dynamic flow algorithms. These are
the tasks of reading and writing to or from the cloud, invoked by a
single node (\CW and \CR), or by all nodes (\CAW
and \CAR).
Our goal in all the tasks and algorithms 
is to find a schedule that implements the task in the minimum amount of time.
\begin{definition}\label{def-sched}
  Given a \CWC model, a \textbf{schedule} for time interval
  $I$ is a mapping that assigns, for each time step in $I$ and each link $(u,v)$: a send
  (or null) operation
  if $(u,v)$ is a local link, and a \FW or \FR (or null) operation if $(u,v)$ is a
  cloud link.
\end{definition}

We  present optimal solutions to these problems in general directed
graphs, using the quickest flow algorithm.
\para{Serving a Single Node:}
\ignore{
	The dynamic flow problem~\cite{Skutella-flows,Ahuja-Flows} is a
	generalization  of the maximum flow
	problem that dates back to Ford and
	Fulkerson \cite{Ford-Fulkerson}.  
	\QstF \cite{Minieka-flow, Wilkinson-flow} is a variant of dynamic
	flow. In this problem 
	there is a graph $G=(V,E)$ with source node $s\in V$ and a target node
	$t\in V$ with each edge $e$ having bandwidth $w(e)$ and traversal time
	$t(e)$.  The goal is to send a given amount of flow units from $s$ to
	$t$ through the network of $G$ in a minimal amount of rounds, without
	exceeding the bandwidth limitations of each edge and with the
	traversal of each edge $e$ requiring $t(e)$ rounds.
	
	The \QstF Problem can be solved using an algorithm for
	the Maximum Flow problem: for a given time estimate $T$, construct a layered graph
	$\FG{G}{T}$ of $T+1$ layers with each layer $i \in \interval{0}{T}$
	containing a copy of all nodes in $V$,
	$\left\{ (v, i) \mid v\in V \right\}$.
	For each node $v$ and layer $i\in \interval{0}{T-1}$, connect $(v,i)$ and $(v,i+1)$
	with a link of infinite bandwidth.
	For each link $e=(u,v)\in E$ and for each $i \in \interval{0}{T-t(e)}$, connect node
	$(u, i)$ to $(v, i+t(e))$ with a link of bandwidth $w(e)$.
	\todo{add a figure?}
}
Let us consider \CW first.
We start with a lemma stating the close relation between schedules (\defref{def-sched})
implementing \CW
in our model and dynamic flows.

\begin{lemma}\label{lem:schedule-iff-dynamic}
	Let $G=(V,E,w)$ be a graph in the \CWC model.
	Then there is a schedule implementing $\CW_i$ from processing node $i$
	to cloud node $v_c$ with message $S$ of size $s$
	in $T$ rounds
	if and only if there is a dynamic flow of value $s$ and time horizon $T$ from source node $i$
	to sink node $v_c$.
\end{lemma}
\Proof
Converting a schedule to a dynamic flow is trivial,
as send and receive operations between processing nodes and \FW and \FR operations
directly translate to a dynamic flow that transports the same amount of flow
while maintaining bandwidth capacities and resulting with a desired dynamic flow.
For the other direction (converting a dynamic flow to a schedule),
let $f$ be a dynamic flow of time horizon $T$ and value $s$ from node $i$ to the cloud $v_c$.
We describe a schedule $A$ implementing $\CW_i$
as follows.

First, we construct another dynamic flow $f'$ which is the same as $f$,
except that no flow leaves the sink node $v_c$.
Formally, let $\val_g(v,t)$
be the total number of flow units that are stored in node $v$ at time step $t\in\Set{0,T}$
according to some dynamic flow $g$.
Flow $f'$ is constructed by induction;
In time step $1$, $f'$ is the same as $f$, except for setting
$f'(e,1) = 0$ for every edge $e$ that leaves $v_c$.
Let $t\in \Set{1,T-1}$. In step $t+1$, $f'$ is defined the same as $f$,
while truncating the sum of all flow that leaves node $v$ to be at most $\val_{f'}(v,t)$
for every node $v$, and setting $f'(e,1) = 0$ for every edge $e$ that leaves the sink.

Let $G_p = G-\{v_c\}$, and let $\val_g(G_p, t)$  denote $\sum_{v\in V_p}\!\val_g(v,t)$
for some dynamic flow $g$.
 Initially, $\val_f(G_p, 0) = \val_{f'}(G_p, 0) = s$,
and that by the induction, in every step $t\in \Set{1,T}$,
$\val_f(G_p, t) \ge \val_{f'}(G_p, t)$ due to flow units not being able to
get to $G_p$ from $v_c$.
Finally, in time step $T$, $\val_f(G_p, T) = 0$ since all flow was sent to the sink,
and thus $\val_{f'}(G_p, T) = 0$ as well
and $\val_{f'}(v_c,T) = s$ due to flow conservation.
Therefore, $f'$ is a dynamic flow of time horizon (at most) $T$ and value $s$.

Now, we use $f'$ to describe the volume of data that is sent along each link
in every round of the schedule $A$, by translating flow between processing
nodes to send/receive operations, and between processing and cloud nodes to \FW operations
(no \FR operations are required according to $f'$, because no flow leaves $v_c$).
To specify which data is sent in every operation of the schedule,
refer to all \FW operations of the schedule.
Let $K$ be the number of \FW operations in $A$, let $i_k$ be the node
initiating the $k$-th call to \FW and let $l_k$ be
the size of the message in that call.
We assign the data transferred on each link during $A$ so that
when node $i_k$ runs the $k$-th \FW, it would write the part of $S$
starting at index
$s\cdot{ \sum_{j=1}^{k-1}{l_j}} $
and extending for $l_k$ bits.
Note that processing nodes do not need to exchange indices of the data they transfer,
as all nodes can calculate in preprocess time the schedule and thus
``know in advance''
the designated indices of the transferred data.

Correctness of the schedule $A$ follows from the validity of $f$, as well as its 
time complexity.
\QED

\begin{theorem}
\label{thm-cw}
Given any instance $G=(V,E,w)$ of the \CWC model, an optimal schedule
realizing $\CW_i$ can be computed in polynomial time.
\end{theorem}
\Proof
Consider a $\CW$ issued by a
processing node $i$, wishing to write $s$ bits to cloud node $v_c$. We
construct an instance of quickest flow as follows.
The flow network is $G$ where $w$ is the link capacity function, node $i$
is the source and $v_c$ is the sink. The requested flow
value is $s$. The solution, computed by \theoremref{thm-quickest},  is directly translatable to a schedule,
after assigning index ranges to flow parts according to \lemmaref{lem:schedule-iff-dynamic}.
Optimality of the resulting schedule follows from the optimality of
the quickest flow algorithm.
\QED
\para{Remarks.}
\begin{compactitem}
	\item Interestingly, in the presence of multiple cloud
	nodes, it may be the case that while writing to one cloud
	node, another cloud node is used as a relay station.
	\item Schedule computation can be carried out
	off-line: we can compute a schedule for each node $i$ and for each
	required file size $s$ (maybe it suffices to consider only powers of $2$),
	so that in run-time,
	the initiating node would only need to tell all other nodes which schedule to use.
	\footnote{
		We note that sending the initiating messages can be problematic if it uses
		cloud nodes as relays, and may require some sort of synchronization across all
		processing nodes.
	}
\end{compactitem}

\medskip\noindent
Finally, we observe that the reduction sketched in the proof of
\theoremref{thm-cw}  works for reading just as well: the only
difference is reversing the roles of source and sink, i.e.,
pushing $s$ flow units from the cloud node $v_c$ to the requesting
node $i$. We therefore have also:
\begin{theorem}
\label{thm-cr}
Given any instance of the \CWC model, an optimal schedule realizing
$\CR_i$ can be computed in polynomial time.
\end{theorem}

\para{Serving Multiple Nodes:}
Consider now operations with multiple invocations. Let us start with
$\CAW$ (\CAR is analogous, as mentioned above). Recall that in this
task, each node has a (possibly empty) file to write to a cloud
node. If all nodes write to 
the same cloud node, then using
the evacuation problem
variant of the
quickest flow algorithm solves the problem (see \defref{def-evacuation}),
However, if we need to
write to multiple cloud nodes, we resort to the quickest \emph{multicommodity}
flow, defined as follows~\cite{Skutella-flows}.

We are given a flow network as described in \sectionref{ssec:dynamic},
but with $k$ source-sink pairs $\Set{(s_i,t_i)}_{i=1}^k$, and $k$
\emph{demands} $d_1,\ldots,d_k$. We seek $k$ flow functions $f_i$,
where $f_i$ describes the flow of $d_i$ units of commodity $i$ from
its source $s_i$ to its sink $t_i$, subject to the usual constraints:
the edge capacity constraints (\ref{const-edge}) applies to the sum of
all $k$ flows, and the node capacity constraints (\ref{const-node}-\ref{const-node-end}),
as well as the source and sink constraints
(\ref{const-source}-\ref{const-sink}) are written for each commodity
separately.

For the case that all nodes write to the same cloud node,
we can get the following theorem.
\begin{theorem}
\label{thm-caw-car}
Given any instance $G=(V,E,w)$ of the \CWC model, an optimal schedule
realizing \CAW (\CAR) in which every node $i$ needs to write (read) a
message of size $s_i$ to (from) cloud node $v_c$ can be computed in
strongly polynomial time.
\end{theorem}
\Proof
Similarly to \lemmaref{lem:schedule-iff-dynamic}, one can see that there is such an algorithm
for \CAW if and only if there is a dynamic flow solving the evacuation problem 
(\defref{def-evacuation}).
Thus in order to solve \CAW, we simply need to construct a flow network as is done
in \theoremref{thm-cw}, and apply \theoremref{thm-evacuation} in order to get the solution,
that can be then translated to a schedule.

A schedule for \CAR can be obtained by reversing the schedule for \CAW,
similarly to \theoremref{thm-cr}.
\QED

\bigskip
For the case of multiple targeted cloud nodes,
it is known that 
determining whether there exists a feasible quickest multicommodity 
flow with a given time horizon $T$
is NP-hard, but on the positive side, there exists an FPTAS to
it~\cite{quickest-multi}, i.e., we can approximate the optimal $T$ to
within $1+\epsilon$, for any constant $\epsilon > 0$.
Extending the reduction for
single commodity in the natural way, we obtain the following result.

\begin{theorem}
\label{thm-caw-car-multi}
Given any instance of the \CWC model and $\epsilon>0$, a schedule
realizing $\CAW$ or $\CAR$ can be computed in time polynomial in the
instance size and $\epsilon^{-1}$. The length of the schedule is at
most $(1+\epsilon)$ times larger than the optimal length.
\end{theorem}

\section{Computing \& Writing Combined Values}
\label{sec:va}
 Flow-based techniques are not applicable in the case
of writing a combined value, because the very essence of combining
violates conservation constraints (i.e., the number of bits entering a node may be different than the number of bits leaving it). However, in
\subsectionref{sec-ccw-flow} we explain how to implement \CCW in the
general case using \CAW and \CAR. While simple and generic, these
implementations can have time complexity much larger than optimal.
We offer partial remedy in \subsectionref{sec-fat-combining}, where we
present our main result: an algorithm for \CCW 
when $\otimes$ is commutative and the local network has ``fat
links,'' i.e., all local links have capacity at least $s$.
For this
important case, we show how to complete the task in
time larger than the optimum by an $O(\log^2n)$ factor.

\subsection{Combining in General Graphs}
\label{sec-ccw-flow}
\begin{algorithm}[t]
  \caption{High-level algorithm for \CCW using \CAW and \CAR}
  \label{alg:caw-flow}
  \begin{algorithmic}[1]\small
    \STATE $m := n$, $j := 0$
    \STATE for all $i<n$ set $X_i^0=S_i$, and for all $i > n$, $X_i^0=\Unit$
    \WHILE {$m > 1$} \label{alg:caw-flow:loop}
    \STATE run \CAW with inputs $S_i=X_i^j$
    \label{alg:caw-flow:caw}
    \STATE run \CAR with inputs $S_i=X_{2i}^j$
    \STATE run \CAR with inputs $S_i=X_{2i+1}^j$
    \STATE $m := \ceil{m/2}$ \label{alg:caw-flow:advance}
    \STATE for all $i<m$ set $X_i^{j+1}=X_{2i}^j \otimes X_{2i+1}^j$,
    and for all $i>m$, $X_i^{j+1}=\Unit$
    \STATE for all $i < m$, in parallel, node $i$ calculates $X_i^{j+1}$ locally
    \STATE $j := j+1$
    \ENDWHILE
    \STATE run \CW from node $0$ to write $X_0^j$ to the cloud
    \label{alg:caw-flow:last-caw}
  \end{algorithmic}
\end{algorithm}

We now present algorithms for \CCW 
and for \CCR on general graphs, using
the primitives treated in \sectionref{sec:rw}. 
%
Note that with a
non-commutative operator, the operands must be ordered; using renaming if necessary, we assume
w.l.o.g.\ that in such cases
the nodes are indexed by the same order of their operands.


\begin{theorem}
	\label{thm-ccw-gen}
	Let $T_s$ be the running time  of \CAW (and \CAR)
	when all files have size $s$.  Then \algref{alg:caw-flow} solves \CCW 
	in $O(T_s \log{n})$ rounds.
\end{theorem}
\begin{figure}[htb]
  \begin{center}
    \begin{tikzpicture}
      \node [draw]{$X_0^3$} [level distance=10mm,sibling distance=50mm]
      child {
        node [draw]{$X_0^2$} [level distance=10mm ,sibling distance=25mm]
        child {
          node [draw] {$X_0^1$} [level distance=10mm ,sibling distance=12mm]
          child {node [draw] {$X_0^0$}}
          child {node [draw]{$X_1^0$}}
        }
        child {
          node [draw]{$X_2^1$} [level distance=10mm ,sibling distance=12mm]
          child {node [draw] {$X_2^0$}}
          child {node [draw] {$X_3^0$}}
        }
      }
      child {
        node [draw]{$X_4^2$} [level distance=10mm ,sibling distance=25mm]
        child {
          node [draw] {$X_4^1$} [level distance=10mm ,sibling distance=12mm]
          child {node [draw] {$X_4^0$}}
          child {node [draw]{$X_5^0$}}
        }
        child {
          node [draw]{$X_6^1$} [level distance=10mm ,sibling distance=12mm]
          child {node [draw] {$X_6^0$}}
          child {node [draw] {$X_7^0$}}
        }
      };
    \end{tikzpicture}
  \end{center}
  \caption{\em Computation tree  with $n=8$.
    $X_i^j$ denotes the result stored in node $i$ in iteration $j$.}
  \label{fig-tree}
\end{figure}
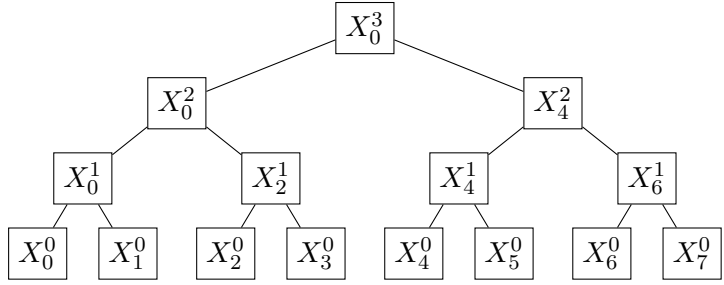

\Proof
By algorithm. The idea is to do the combining over a binary
``computation tree'' by
using the cloud to store the partial results.
The computation tree is defined as follows (see
\figref{fig-tree}).
Let $X_i^j$ denote the $i$-th node at level $j$, as well as the
value of that node.
The leaves $X_i^0$ are the input values, and the value of
an internal node $X_{i}^{j+1}$ at level $j+1$ with left child
$X_{2i}^{j}$ and right child $X_{2i+1}^{j}$ is $X_{2i}^{j}\otimes
X_{2i+1}^{j}$. Pseudocode is provided in
\algref{alg:caw-flow}. Correctness of the algorithm follows from the
observation that after each execution of Step \ref{alg:caw-flow:caw},
there are 
$m$ files of size $s$ written in the cloud, whose product is the
required output, and that $m$ is halved in every iteration.
If at any iteration $m$ is odd,
then node $\ceil{m/2}-1$ only needs to read one file, and therefore we
set the other file that it reads to be $\Unit$.
When $m$ reaches $1$, there is only $1$ file left, which is the
required result. 


As for the time analysis: Clearly, a single iteration of the \textbf{while} loop 
takes $3T_s = O(T_s)$ rounds.
Note that when writing $\Unit$ to the cloud, it can be encoded as a $0$-bit string,
which can only improve the runtime.
There are $\ceil{\log n}$ rounds due to Step \ref{alg:caw-flow:advance},
and Step \ref{alg:caw-flow:last-caw} is also completed in $O(T_s)$ rounds,
and thus the total runtime of the algorithm is $O(T_s \log{n})$.
\QED

In a way, \CCR is the ``reverse'' problem of \CCW, since it starts
with $s$ bits in the cloud and ends with $s$ bits of output
in every node. However, \CCR is easier than \CCW because our model allows for
concurrent reads and disallows concurrent writes.
We have the following result.

\begin{theorem}
	\label{thm-ccr-gen}
	Let $T_s$ be the time required to solve \CAR 
	when all files have size $s$.  Then \CCR can be solved in $T_s$ rounds as
	well. 
\end{theorem}
\Proof 
First note that if there were $n$ copies of the input file
$S$ in the cloud, then \CCR and \CAR would have been the exact same
problem.  The theorem follows from the observation that any algorithm
for \CAR with $n$ inputs of size 
$s$ in the cloud can be modified so that each invocation of $\FR$ with
argument $S_i$ is converted to \FR with argument $S$ (the input of
\CCR).
\QED


\subsection{Combining Commutative Operators in Fat links Network}
\label{sec-fat-combining}
In the case of $s$-fat-links network (i.e.,  all local links are
have
bandwidth at least 
$s$, and all links are symmetric) we
can construct a near-optimal algorithm for \CCW.
The idea is to use multiple $\CW$ and \CR operations
instead of \CAW and \CAR. The challenge is to
minimize the number of concurrent operations per node;
to this end we use 
sparse covers \cite{AP90}.

We note that if the network is $s$-fat-links but the operand size is
$s'>s$, the algorithms still apply, with an additional factor of
$\ceil{s'/s}$ to the running time.
The lower bounds in this \dsubsection, however, may change
by more than that factor.

We start with a tight analysis of \CW and \CR in this setting and then
generalize to \CCW and \CCR.

\paragraph{Implementation of \CW and \CR.}
Consider $\CW_i$, where $i$ wishes to write $s$ bits to a given cloud node. 
The basic tension in finding an optimal schedule for $\CW_i$ is that in order to use more cloud bandwidth, more nodes need to be enlisted. But while more bandwidth reduces the transmission
time, reaching remote nodes (that provide the extra bandwidth)
increases the traversal time.
Our algorithm looks for the
sweet spot where the conflicting effects are more-or-less balanced.

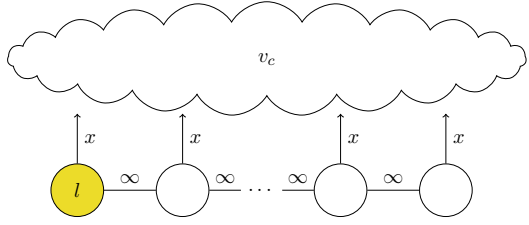
\begin{figure}[t]
	\centering
	\scalebox{.7}{
		\begin{tikzpicture}
			\node[draw, circle, inner color=white, minimum size=1cm] at (0,0) (mid) {$$};
			\node[draw, circle, fill=yellow!90!black, minimum size=1cm] at (-2,0) (l) {$l$};
			\node[draw, circle, inner color=white, minimum size=1cm] at (3,0) (r) {};
			\node[draw, circle, inner color=white, minimum size=1cm] at (5,0) (rfar) {$$};
			\node at ($(mid)!.5!(r)$) (dots) {\ldots};
			
			\node [cloud, draw,cloud puffs=23,cloud puff arc=120, aspect=7, inner ysep=1em]
			at 	(1.6, 2.5)
			{\emph{$v_c$}};
			
			\draw [-] (mid) -- (l)
			node [above,text width=3cm,text centered,midway] {$\infty$};
			\draw [-] (mid) -- (dots)
			node [above,text width=3cm,text centered,midway] {$\infty$};
			\draw [-] (r) -- (dots)
			node [above,text width=3cm,text centered,midway] {$\infty$};
			\draw [-] (r) -- (rfar)
			node [above,text width=3cm,text centered,midway] {$\infty$};
			\draw [->] (mid) -- (0, 1.45)
			node [right,text width=0cm,text centered,midway] {$x$};
			\draw [->] (r) -- (3, 1.45)
			node [right,text width=0cm,text centered,midway] {$x$};
			\draw [->] (l) -- (-2, 1.45)
			node [right,text width=0cm,text centered,midway] {$x$};
			\draw [->] (rfar) -- (5, 1.45)
			node [right,text width=0cm,text centered,midway] {$x$};
		\end{tikzpicture}
	}
	\caption{ \em
		A simple path example. The optimal distance to travel in order to
		write an $s$-bits file to the cloud would be $\sqrt{s/x}$.
	}
	\label{fig-tradeoff-example}
\end{figure}

For example, consider a simple path of $n$ nodes with infinite local
bandwidth, where each node is connected to the cloud with bandwidth
$x$ (\figref{fig-tradeoff-example}).  Suppose that the leftmost node
$l$ needs to write a message of $s$ bits to the cloud.  By itself,
writing requires $s/x$ rounds.  Using all $n$ nodes, uploading would
take $O(s/nx)$ rounds, but $n-1$ rounds are needed to ship the
messages to the fellow-nodes.  The optimal solution in this case is to
use only $\sqrt{s/x}$ nodes: the time to ship the file to all these
nodes is $\sqrt{s/x}$, and the upload time is
$\frac{s/\sqrt{s/x}}{x} = \sqrt{s/x}$, because each node needs to
upload only $s/\sqrt{s/x}$ bits.

In general, we define 
``cloud clusters'' to be
node sets that optimize the ratio between their diameter
and their total bandwidth to the cloud.
Our algorithms for \CW and \CR use nodes of
cloud clusters.
We prove that the running-time of our implementation is
asymptotically optimal.
Formally, we have the following.

\begin{definition}\label{def-clusters}
	Let $G=(V,E,w)$ be a \CWC system with processor nodes $V_p$ and cloud nodes $V_c$.
	The \textbf{cloud bandwidth} of a processing node $i \in V_p$ w.r.t.\ a
	given cloud node $v_c\in V_c$ is $\bc(i)\DEF w(i,v_c)$.
	A \textbf{cluster} $B \subseteq V_p$ in $G$ is a connected set
	of processing nodes. The \textbf{cloud (up or down) bandwidth} of cluster $B$ w.r.t\ a given
	cloud node,
	denoted $\bc(B)$, is the sum of the cloud bandwidth to $v_c$ over all nodes in
	$B$: $\bc(B)\DEF\sum_{i\in B}\bc(i)$.  The (strong) \textbf{diameter}
	of cluster $B$, denoted $\Diam(B)$, is the maximum distance between
	any two nodes of $B$ in the induced graph $G[B]$:
	$\Diam(B) = \max_{u,v \in B} {dist_{G[B]}(u, v)}$.
\end{definition}

We use the following definition for the network when ignoring the
cloud.
Note that the metric here is hop-based---$w$
indicates link bandwidths.

\begin{definition}\label{def-Ball}
	Let $G=(V,E,w)$ be a \CWC system with processing nodes $V_p$ and
	cloud nodes $V_c$.  The \textbf{ball of radius $r$
		around node $i\in V_p$}, denoted $\Br r i$ is the set of nodes at most $r$ hops away
	from $i$ in $G_p$.
\end{definition}

Finally, we define the concept of cloud cluster of a node.

\begin{definition}\label{def-Bi}
	Let $G=(V,E,w)$ be a \CWC system with processing nodes $V_p$ and
	cloud node $v_c$, and let $i\in V_p$.
	Given $s\in\mathbb{N}$,  the \textbf{$s$-cloud radius} of
	node $i$, denoted $\kkR{i}s$, is defined to be
	$$
        \kkR{i}s
        \DEF
        \min( \Set{\Diam(G_p)}
                 \,\cup\,
              \Set{k\mid  
		(k\tp 1) \cdot \bc(\Br{k}{i}) \ge s})~.
	$$
	The ball $\BR{i}\DEF \Br{\kkR{i}s}{i}$ is the $s$-\textbf{cloud
		cluster} of node $i$.
	The \textbf{timespan} of the $s$-cloud cluster of $i$
	is denoted $\ZR{i}\DEF \kkR{i}{s} + \frac{s}{\bc(\BR{i})}$.
	We sometimes omit the 
	$s$ qualifier when it is clear from the context.
\end{definition}
In words, $\BR{i}$ is a cluster of  radius $\kR i $ around node $i$,
where $\kR i$ is the smallest radius that allows writing $s$ bits to $v_c$ by
using all cloud bandwidth emanating from
$\BR i$ for $\kR i+1$ rounds.
$\ZR{i}$ is the time required  (1) to send $s$ bits from node $i$ to all nodes in
$\BR{i}$, and (2) to upload $s$ bits to $v_c$ collectively by 
 all  nodes of $\BR{i}$.
Note that $\BR{i}$ is easy to compute.  We can now state our upper bound.
\begin{theorem}\label{theorem:CW_UB}
	Given a fat-links \CWC system, \algref{alg:cw-hi} solves the
	$s$-bits $\CW_i$ problem in $O(\ZR{i})$ rounds on $\BR{i}$.
\end{theorem}

\Proof 
The algorithm broadcasts all $s$ bits to all nodes in $\BR i$,
and then each node writes a subrange of the data whose size is
proportional to its cloud bandwidth. Correctness is obvious.
As for the time analysis:
Steps \ref{cw-alg-BFS-tree}-\ref{cw-alg-send-local} require $O(\kR i)$ rounds.
In the loop of steps \ref{cw-alg-node-write}-\ref{cw-alg-ack},
$\bc(\BR i)$ bits are sent in every round, and thus it terminates in
$O({s}/{\bc(\BR i)})$ rounds. The theorem follows from the definition of $\ZR i$.
\QED

\begin{algorithm}[t]\small
  \caption{$\CW_i$ }
  \label{alg:cw-hi}
  \begin{algorithmic}[1]
    \STATE \label{cw-alg-BFS-tree}
    Construct a BFS spanning tree of $\BR{i}$ rooted at node $i$
    and assign for each index $1 \le x \le |\BR{i}|$ a unique node
    $v(x)\in \BR{i}$ according to their BFS order ($v(1)=i$)
    \STATE \label{cw-alg-send-local}
    Broadcast $S$ from node $i$ to all nodes in $\BR{i}$ using the tree
    \FOR{all $x:=1$ \textbf{to} $|\BR{i}|$, in parallel}
    \STATE\label{cw-alg-node-write}{
      Node $v(x)$ writes to the cloud the part of $S$ starting at
      $s\cdot\frac{ \sum_{y=1}^{x-1}{\bC{v(y)}} }{\bC{\BR i}} $
      and extending for $s\cdot\frac{\bC{v(x)}}{\bC{\BR i}}$ bits,
      writing $\bC{v(x)}$ bits in every round.}
    \STATE\label{cw-alg-ack}{Node $v(x)\ne i$ sends an acknowledgment to $i$
      when done, and halts}
    \ENDFOR    
    \STATE\label{cw-alg-end} Node $i$ halts when all acknowledgments
    are received.
    \hfill{\em // for \CR reversal}
  \end{algorithmic}
\end{algorithm}

\smallskip
Next, we show that our solution for $\CW_i$ is optimal, up to a
constant factor.  We consider the case of an incompressible input
string: such a string exists for any size $s\in\mathbb{N}$ (see, e.g.,
\cite{LiV:Kolmogorov}).  As a consequence, {in any execution of a
	correct algorithm, $s$ bits must cross any cut that separates $i$
	from the cloud node}, giving rise to the following 
	lower bound.

\begin{theorem}
  \label{theorem:CW_LB}
  Any algorithm solving $\CW_i$ in a fat-links \CWC requires $\Omega(\ZR{i})$ rounds.
\end{theorem}

\Proof 
By definition, $\ZR{i} = \kR{i} + {s}/{\bc(\BR{i})}$.
\lemmaref{lemma:cluster_diameter} and \lemmaref{lemma-cluster-bc} show
that each term of $\ZR{i}$ is a lower bound on the running time of any
algorithm for $\CW_i$.  \QED

\begin{lemma}
  \label{lemma:cluster_diameter}
  Any algorithm solving $\CW_i$ in a fat-links \CWC system requires
  $\Omega \left( \kR{i}\right)$ rounds.
\end{lemma}

\Proof Let $A$ be an algorithm for $\CW_i$ that writes string $S$  in
$t_A$ rounds. 
If $t_A \ge \Diam(G_p)$ then $t_A \ge \kR{i}$ and we are
done. Otherwise, we count the number of bits of $S$ that can get to
the cloud in $t_A$ rounds.  Since $S$ is initially stored in $i$, 
in a given round $t$, only nodes in $\Br{t-1}{i}$ can write pieces of
$S$ to the cloud. 
Therefore, overall, $A$ writes to the cloud at most
$\sum_{t=1}^{t_A}\bc(\Br{t-1}{i}) \le t_A\cdot\bc(\Br{t_A-1}{i})$
bits. Hence, by assumption that $A$ solves $\CW_i$, we must have
$t_A\cdot\bc(\Br{t_A-1}{i}) \ge s$. The lemma now follows from
definition of $\kR{i}$ as
the minimal integer
%
$\ell \le \Diam(G_p)$ such that
$(\ell \tp 1) \cdot \bc( \Br{\ell}{i} ) \ge s$ if it exists
(otherwise, $\kCR{i}=\Diam(G_p)$ and $t_A>\Diam(G_p)$). Either way,
we are done.
\QED

\begin{lemma}\label{lemma-cluster-bc}
  Any algorithm solving $\CW_i$ in a fat-links \CWC system requires
  $\Omega \left({s}/{\bC{\BR{i}}} \right)$ rounds.
\end{lemma}

\Proof Let $A$ be an algorithm that solves $\CW_i$ in $t_A$ rounds.
If $\kR{i}=\Diam(G_p)$ then $\BR{i}$ contains all processing nodes
$V_p$, and the claim is obvious, as no more than $\bC{V_p}$ bits can
be written to the cloud in a single round.  Otherwise,
$\kR{i} \ge {s}/{\bC{\BR{i}}} - 1$ by \defref{def-Bi}, and we are done
since $t_A = \Omega(\kR{i})$ by \lemmaref{lemma:cluster_diameter}.
\QED

By reversing time (and hence information flow) in a schedule of \CW, one gets a schedule for \CR. 
Hence we have the following immediate corollaries.

\begin{theorem}\label{theorem:CR_UB}
	$\CR_i$ can be executed in
	$O(\ZR{i})$ rounds in a fat-links \CWC.
\end{theorem}


\begin{theorem} \label{theorem:CR_LB} 
	$\CR_i$ in a fat-links \CWC    requires $\Omega(\ZR{i})$ rounds.
\end{theorem}

\para{Remark:}
The lower bound of \theoremref{theorem:CW_LB}
and the definition of {cloud clusters} (\defref{def-Bi})
show an interplay between the
message size $s$, cloud bandwidth, and the network diameter;
For large enough $s$, the cloud cluster of a node includes all processing nodes (because the time spent crossing the local network is negligible relative to the upload time), and for small enough $s$, the cloud cluster includes only
the invoking node, rendering the local network redundant.

\paragraph{Implementation of \CCW.}
\label{ssec:combining}
%
Below, we first show how 
to implement \CCW using any given cover.
In fact, we shall use \emph{sparse covers}~\cite{AP90},
which allow us to to get near-optimal performance.

Intuitively, every node $i$ has a cloud cluster $\BR{i}$ which allows it
to perform $\CW_i$, and calculating the combined value within every cloud cluster $\BR{i}$
is straight-forward (cf.\
\algref{alg:ccw-low} and \lemmaref{lemma:ccw-low-runtime}).
Therefore, given a partition of the graph that consists of pairwise disjoint
cloud clusters, \CCW can be solved by
combining the inputs in every cloud cluster, followed by
combining the partial results in a computation-tree fashion using
\CW and \CR.
However, such a partition may not always exist, and we resort
to a \emph{cover} of the nodes.
Given a cover $\CC$ in which every node is a member of at most $\load(\CC)$ clusters,
we can use the same technique, while increasing the running-time by a factor of $\load(\CC)$
by time multiplexing.
Using Awerbuch and Peleg's sparse covers (see \theoremref{thm-AP}),
we can use an initial cover $\CC$ that consists of all cloud clusters in the graph
to construct another cover, $\CC'$, in which $\load(\CC')$ is  $O(\log n)$,
paying an $O(\log n)$ factor in cluster diameters,
and use $\CC'$ to get near-optimal results.


%

\begin{definition}\label{def-clusters-params}
	Let $G$ be a \CWC system, and let $B$ be a cluster in $G$ (see
	\defref{def-clusters}).  The \textbf{timespan of node $i$ in $B$},
	denoted $Z_B(i)$, is the minimum number of rounds required to
	perform $\CW_i$ (or $\CR_i$), using only nodes in $B$.
	The \textbf{timespan of cluster} $B$, denoted $Z(B)$, is given by
	$Z(B) = \min_{i\in B}Z_B(i)$.  The \textbf{leader} of cluster $B$,
	denoted $r(B)$, is a node with minimal timespan in $B$,
	i.e., $r(B) = \argmin_{i\in B} Z_B(i)$.
\end{definition}
In words, the timespan of cluster $B$ is the minimum time required
for any node in $B$
to write an $s$-bit
string to the cloud using only nodes of $B$.

From \theoremref{theorem:CW_UB}, we get the following lemma:
\begin{lemma} \label{lemma:cloud-cluster-timespan}
	For every node $i$, $Z(\BR{i}) \le Z_{\BR{i}}(i) = O(Z_i)$.
\end{lemma}

\begin{definition}
	\label{def-covers}
	Let $G$ be a \CWC system with processing node set $V_p$.
	A \textbf{cover} of $G$ is a set of clusters
	$\CC = \{ B_1, \ldots, B_m \}$ such that 
	$\cup_{B \in \cC} B = V_p$.
	The \textbf{load of node $i$} in a cover $\CC$ is the number of
	clusters in $\CC$ that contain $i$, i.e.,
	$\load_{\CC}(i) = |\{B \in \CC \mid i \in B\}|$.
	The \textbf{load of cover $\CC$} is the maximum load of any node
	in the cover, i.e.. $\load(\CC) = \max_{i\in V_p}
	{\load_{\CC}(i)} $.  
	The \textbf{timespan} of cover $\CC$, denoted
	$\ZC(\CC)$, is the maximum timespan of any cluster in $\CC$,
	$\ZC(\CC) = \max_{B\in \CC} {Z(B)}$.
	The \textbf{diameter} of cover $\CC$, denoted $\Diam_{\max}(\CC)$,
	is the maximum diameter of any cluster in $\CC$,
	$\Diam_{\max}(\CC) = \max_{B\in \CC} {\Diam(B)}$.
\end{definition}

\begin{algorithm}[t]
	\caption{High-level algorithm for \CCW given a cover $\CC$}
	\label{alg:ccw-hi}
	\begin{algorithmic}[1]
		\STATE\label{cluster-graph}
		For a node $i\in V_p$, let $C_i[1],C_i[2],\ldots$ be the clusters containing $i$.
		
		\STATE\label{ccw-1}
		For the rest of the algorithm, multiplex each round as $\load(\CC)$ rounds,
		such that each node $i$ operates in the context of cluster $C_i[j]$ 
		in the $j$-th round .
		\FORALL{$B\in \CC$, in parallel}
		\label{ccw-startloop}
		\STATE\label{ccw-local}{
			Compute $P_B=\bigotimes_{j\in B}S_j$ using \algref{alg:ccw-low}
			\hfill{\em // convergecast using local links only}
		}
		\ENDFOR
		\label{ccw-endloop}
		\STATE\label{ccw-high} 
		Apply 
		\algref{alg:ccw-high:node}
		\hfill{\em // the result is stored in the cloud}
	\end{algorithmic}
\end{algorithm}

We now give an upper bound in terms of any given cover $\CC$.
\begin{theorem}\label{thm-ccw-covers-ub}
	\algref{alg:ccw-hi} solves \CCW in a
	fat-links \CWC in	$O\left(\Diam_{\max}(\CC) \cdot \load(\CC) + \ZC(\CC) \cdot \load(\CC) \cdot \log |\CC|\right)$
	rounds. 
\end{theorem}

The basic strategy is to first compute the combined value 
in each cluster 
using only the local links,
and then combine the cluster values using a
computation tree. However, unlike \algref{alg:caw-flow}, we use
\CW and \CR 
instead of \CAW and \CAR.

A high-level description is given in \algref{alg:ccw-hi}.  The
algorithm consists of a preprocessing part (lines
\ref{cluster-graph}-\ref{ccw-1}), and the execution part, which consists
of the ``low-level'' computation using only local links (lines
\ref{ccw-startloop}-\ref{ccw-endloop}), and the ``high-level''
computation among clusters (line \ref{ccw-high}).
We elaborate on each below.

\para{Preprocessing.}
A major component of the preprocessing stage is computing the cover $\cC$, which we specify later (see \theoremref{thm-ccw-ub}). In \algref{alg:ccw-hi}
we describe the algorithm as if it operates in each cluster independently of
other clusters, but clusters may overlap. To facilitate this mode of operation, we use time multiplexing: nodes execute work on behalf of the clusters they are member of in a round-robin fashion, as specified 
in lines \ref{cluster-graph}-\ref{ccw-1} of \algref{alg:ccw-hi}.
This allows us to
invoke operations limited to clusters in all clusters
``simultaneously'' 
by increasing the time complexity by a
$\load(\CC)$ factor.


\para{Low levels: Combining within a single cluster.}
To implement line \ref{ccw-local} of \algref{alg:ccw-hi}, we build,
in each cluster $B\in \CC$, a
spanning tree rooted at $r(B)$, and apply
Convergecast~\cite{Peleg:book} using $\otimes$.  
Ignoring the multiplexing of \algref{alg:ccw-hi}, we have:
\begin{lemma} \label{lemma:ccw-low-runtime} \algref{alg:ccw-low}
	computes $P_B = \bigotimes_{i\in B} S_i$ at node $r(B)$  in   $O(\Diam(B))$ rounds.
\end{lemma}
 To get the right overall result, each input $S_i$
is associated with a single cluster in $\CC$.  To this end, we
require each node to select a single cluster in which it is a member
as its \emph{home cluster}.  When applying \algref{alg:ccw-low}, we
use the rule that the input of node $i$ in a cluster $B\ni i$ is $S_i$
if $B$ is $i$'s home cluster, and $\Unit$ otherwise.




Considering the scheduling obtained by Step \ref{ccw-1},
we get the following lemma.
\begin{lemma} \label{ccw-lem-lo} Steps
	\ref{ccw-startloop}--\ref{ccw-endloop} 
	of \algref{alg:ccw-hi} terminate in
	$O(\Diam_{\max}(\CC) \cdot \load(\CC))$ rounds, with $P_B$ stored
	at the leader node of $B$ for each cluster $B\in C$.
\end{lemma}

\begin{algorithm}[t]
	\caption{Computing the combined result of cluster $B$ at leader $r(B)$}
	\label{alg:ccw-low}
	\begin{algorithmic}[1]
		\STATE 
		Construct a BFS tree of $B$ rooted at node $r(B)$.
		Let $h$ be the height of the tree.
		\FOR{$d:=h$ \textbf{to} $2$} \label{alg:ccw-low:for-loop}
		\FORALL{$i \in B$ at layer $d$ of the tree, in parallel}
		\IF {$i$ is not a leaf}
		\STATE {$i$ computes $S'_i:=S_i \otimes \bigotimes_{j\in child(i)}S'_j$}
		\ELSE
		\STATE {$S'_i:=S_i$}
		\ENDIF
		\STATE {$i$ sends $S'_i$ to its parent node in the tree}
		\ENDFOR
		\ENDFOR    
		\STATE\label{alg:ccw-low:end} Node $r(B)$ computes
		$P_B:=  S_{r(B)} \otimes \bigotimes_{j\in child(r(B))}S'_j$
	\end{algorithmic}
\end{algorithm}

\ignore
{
\begin{figure}[b]
  \begin{center}
    \begin{tikzpicture}[scale=0.8]
      \node [draw]{$X_0^3$} [level distance=10mm,sibling distance=50mm]
      child {
        node [draw]{$X_0^2$} [level distance=10mm ,sibling distance=25mm]
        child {
          node [draw] {$X_0^1$} [level distance=10mm ,sibling distance=12mm]
          child {node [draw] {$X_0^0$}}
          child {node [draw]{$X_1^0$}}
        }
        child {
          node [draw]{$X_2^1$} [level distance=10mm ,sibling distance=12mm]
          child {node [draw] {$X_2^0$}}
          child {node [draw] {$X_3^0$}}
        }
      }
      child {
        node [draw]{$X_4^2$} [level distance=10mm ,sibling distance=25mm]
        child {
          node [draw] {$X_4^1$} [level distance=10mm ,sibling distance=12mm]
          child {node [draw] {$X_4^0$}}
          child {node [draw]{$X_5^0$}}
        }
        child {
          node [draw]{$X_6^1$} [level distance=10mm ,sibling distance=12mm]
          child {node [draw] {$X_6^0$}}
          child {node [draw] {$X_7^0$}}
        }
      };
    \end{tikzpicture}
  \end{center}
  \caption{\em Computation tree  example.
    $X_i^j$ denotes the result stored in $i$ after iteration $j$.}
  \label{fig-tree}
\end{figure}
}

\para{High levels:  Combining using the cloud.}
When \algref{alg:ccw-hi} reaches Step \ref{ccw-high}, the combined
result of every cluster is stored in the leader of the cluster.
The idea is now to fill in a computation tree whose 
leaves are these values
(see~\figref{fig-tree}).

We combine the partial results by filling in the values of a
computation tree defined over the clusters.  The leaves of the tree
are the combined values of the clusters of $\CC$, as
computed by 
\algref{alg:ccw-low}.  To fill in the values of other nodes
in the computation tree, we use the clusters of $\CC$: Each node in
the tree is assigned a cluster which computes its value using the \CR
and \CW primitives.

\begin{algorithm}[t]
	\caption{Computing the high level tree-nodes values}
	\label{alg:ccw-high:node}
	\begin{algorithmic}[1]
		\FOR{$l:=\ceil{\log |C_1|}$ \textbf{to} $1$} 
		\FORALL{tree-nodes $y$ in layer $l$ of the computation tree, in parallel}
		\STATE Let $B := \Clus(y)$
		\IF {$y$ is not a leaf}
		\STATE Let $y_\ell$ and $y_r$ be the left and the right children
		of $y$, respectively.
		\STATE \label{alg:ccw-high:node:read1}
		$r(B)$ invokes \CR for $\Val(y_\ell)$
		\STATE \label{alg:ccw-high:node:read2}
		$r(B)$ invokes \CR for $\Val(y_r)$
		\STATE $r(B)$ computes $\Val(y):=\Val(y_\ell)\otimes \Val(y_r)$
		\ELSE
		\STATE $\Val(y):=P_B$
		\hfill{\em // if $y$ is a leaf its value is already stored at $r(B)$}
		\ENDIF
		\hfill
		\STATE \label{alg:ccw-high:node:write}
		$r(B)$ invokes $\CW$ for $\Val(y)$
		\ENDFOR
		\ENDFOR
	\end{algorithmic}
\end{algorithm}    

Specifically, in \algref{alg:ccw-high:node} we consider a binary tree
with $|\CC|$ leaves, 
where each non-leaf node
has exactly two children.  The tree is constructed from a
complete binary tree with $2^{\ceil{\log|\CC|}}$ leaves,
after deleting the  rightmost $2^{\ceil{\log|\CC|}}-|\CC|$ leaves.
(If by the end the rightmost leaf is the only child of its parent, we delete
the rightmost leaf repeatedly until this is not the case.)


We associate each node $y$ in the computation tree with
a cluster $\Clus(y)\in\cC$ and a value $\Val(y)$, computed by the
processors in $\Clus(y)$ are responsible to compute $\Val(y)$.
Clusters are  assigned to leaves by index:
The $i$-th leaf from the left is associated with the $i$-th cluster of $\CC$.
For internal nodes, we assign the clusters arbitrarily except that we
ensure that no
cluster is assigned to more than one internal node. (This is possible
because in a tree where every node has two or no children, the number
of internal nodes is smaller than the number of leaves.)

The clusters assigned to tree nodes compute the values as
follows (see \algref{alg:ccw-high:node}).
The value associated with a leaf $y_{B}$ corresponding to cluster $B$ is
$\Val(y_{B})=P_B$.
This way, every
leaf $x$ has $\Val(x)$, stored in the leader of $\Clus(x)$,
which can write it to the cloud using \CW.
For an internal node $y$ 
with children $y_l$ and $y_r$,  the leader of
$\Clus(y)$ obtains $\Val(y_l)$ and $\Val(y_r)$ using \CR, computes
their product $\Val(y)=\Val(y_l)\otimes\Val(y_r)$ and invokes $\CW$ to
write it to the cloud. 
The executions of \CW and \CR in a cluster $B$ are done
by the processing nodes of $B$.

Computation tree values are 
filled layer by layer, bottom up.
With this strategy, we have the following result.

\begin{lemma} \label{lemma:alg:ccw-high:node}
	\algref{alg:ccw-high:node} computes $\bigotimes_{i=1}^m{P_i}$
	in $O(\ZC(\CC) \cdot \log{|\CC|})$ rounds,
	assuming that all clusters operate in complete parallelism.
\end{lemma}
\Proof Computing all values in a tree layer requires a constant number
of \CW and \CR invocations in a cluster, i.e., by
\defref{def-covers},
at most $O(\ZC(\CC))$ rounds of work in every layer.
The number of layers is  $\ceil{\log|\CC|}$.
the result follows.
\QED

From 
\lemmaref{lemma:alg:ccw-high:node},
considering the multiplexing of Step \ref{ccw-1},
we get the following result.

\begin{lemma}\label{ccw-lem-hi}
	Line \ref{ccw-high} of \algref{alg:ccw-hi} is completed in
	$O(\ZC(\CC) \cdot \load(\CC) \cdot \log |\CC|)$ rounds.
\end{lemma}
We can therefore conclude:
\par\smallskip\noindent\textbf{Proof of \theoremref{thm-ccw-covers-ub}:~} By \lemmaref{ccw-lem-lo} and
\lemmaref{ccw-lem-hi}. 
\QED

\para{Remark.}
We note that in \algref{alg:ccw-high:node},
Lines \ref{alg:ccw-high:node:read1}, \ref{alg:ccw-high:node:read2} 
and \ref{alg:ccw-high:node:write} essentially compute \CAR and \CAW
in which only
the relevant cluster leaders have inputs.
Therefore, these calls
can be replaced with a collective
call for appropriate \CAR and \CAW,
making the multiplexing of Line \ref{ccw-1} of \algref{alg:ccw-hi} unnecessary
(similarly to \algref{alg:caw-flow}).
By using  optimal schedules for \CAW and \CAR,
the running-time
can only improve beyond the upper bound of \theoremref{thm-ccw-covers-ub}.

\paragraph{Sparse Covers.}
We now arrive at our main result, derived from
\theoremref{thm-ccw-covers-ub}
using a particular flavor of covers.
The result is stated in terms of the maximal timespan of a graph, according to the following
definition.

\begin{definition} \label{graph-Zmax}
	Let $G=(V,E,w)$ be a $CWC$ system with fat links.
	$\Zmax \DEF \max_{i\in V_p} \ZR{i}$ is the \textbf{maximal timespan} in $G$.
\end{definition}
In words, $\Zmax$ is the maximal amount of rounds that is required for any
node in $G$ to write an $s$-bit message to the cloud,
up to a constant factor
(cf.\ \theoremref{theorem:CW_LB}).

\begin{theorem}\label{thm-ccw-ub}
	Let $G=(V,E,w)$ be a \CWC system with fat links.
	Then \CCW with  a commutative combining operator can be solved in $O(\Zmax \log^2 n)$
	rounds.
\end{theorem}

To prove \theoremref{thm-ccw-ub} we use sparse covers. We state
the result from \cite{AP90}.


\begin{theorem}[\cite{AP90}]
	\label{thm-AP}
	Given any cover $\CC$ and an integer $\kappa\ge 1$, a cover
	$\CC'$ that satisfies the following properties can be
	constructed in polynomial time.
	\begin{compactenum}[(i)]
		\item \label{sparse:1} For every cluster $B\in \CC$ there exists a
		cluster $B'\in \CC'$ such that $B \subseteq B'$.
		\item \label{sparse:2}
		$\max_{B'\in \CC'} \Diam(B') \le 4\kappa \max_{B \in \CC}
		\Diam(B)$
		\item \label{sparse:3}
		$\load(\CC') \le 2\kappa |\CC|^{1/\kappa}$.
	\end{compactenum}
\end{theorem}


%
%
\par\noindent\textbf{Proof of \theoremref{thm-ccw-ub}:~}
Let $\CC$ be the cover defined as the set of all cloud
clusters in the system.
By applying \theoremref{thm-AP} to $\CC$ with
$\kappa = \ceil{\log n}$,
we obtain a cover $\CC'$
with $\load(\CC') \le 4\ceil{\log n}$ because $|\CC|\le n$.
By \ref{sparse:2}, $\Diam_{\max}(\CC') \le 4\ceil{\log n} \cdot \Diam_{\max}(\CC)$.
%
Now, 
let $B' \in \CC'$.
We can assume \WLOG that there is a cluster $B\in \CC$ such that $B\subseteq B'$
(otherwise $B'$ can be removed from $\CC'$).
$B$ is a cloud cluster of some node $i\in B'$,
and therefore by \lemmaref{lemma:cloud-cluster-timespan} and by
\defref{def-clusters-params}, we get that $Z(B') \le Z(B) = O(\ZR{i}) = O(\Zmax)$.
Since this bound holds for all clusters of $\CC'$,
$\ZC(\CC') = O(\Zmax)$.

An $
O\left(\Diam_{\max}(\CC) \cdot \log^2 n + \Zmax \cdot \log^2 n\right)$
time bound for \CCW is derived by applying
\theoremref{thm-ccw-covers-ub}
to cover $\CC'$.
Finally, let $B_{j} \in \CC$ be a cloud cluster of diameter $\Diam_{\max}(\CC)$.
Recall that by \defref{def-Bi},
$\Diam(B_{j}) \le 2\kR{j} \le  2\ZR{j} \le 2\Zmax$.
We therefore obtain an upper bound of $O(\Zmax \log^2 n)$ rounds.
\QED


\smallskip
We close with a lower bound.
%
\begin{theorem}\label{theorem:CCW-LB}
	Let $G=(V,E,w)$ be a \CWC system with fat links.
	Then \CCW requires $\Omega(Z_{\max})$ rounds.
\end{theorem}

\Proof 
By reduction from \CW.
Let $i$ be any processing node.
Given $S$, assign $S_i=S$ as the input of node $i$ in \CCW, and for every 
other node $j \ne i$, assign $S_j = \Unit$. Clearly, any algorithm for \CCW
that runs with these inputs solves $\CW_i$ with input $S$.
The result follows from \theoremref{theorem:CW_LB}.
\QED


\paragraph{\CCR.}
\label{sec:CCR}
To implement \CCR, one can reverse the schedule 
of 
\CCW. However, a
slightly better implementation is possible, because there is no need
to ever write to the cloud node.  More specifically, let $\mathcal{C}$
be a cover of $V_p$.  In the algorithm for \CCR, each cluster leader
invokes \CR, and then the leader disseminates the result to all
cluster members.  The time complexity for a single cluster $B$ is
$O(Z(B))$ for the \CR operation, and $O(\Diam(B))$ rounds
for the dissemination of $S$ throughout $B$
(similarly to \lemmaref{lemma:ccw-low-runtime}).
Using the
multiplexing to $\load(\CC)$ as in in Step
\ref{ccw-1} of \algref{alg:ccw-hi}, we obtain the
following result.

\begin{theorem}\label{thm-CCR-covers}
  Let $G=(V,E,w)$ be a \CWC system with fat links.
  Then \CCR can be performed in
  $O(\Zmax \cdot \log^2 n)$ rounds.
\end{theorem}


Finally, we note that since any algorithm for \CCR 
 also solves $\CR_i$ problem for every node $i$, we get from
\theoremref{theorem:CR_LB} the following result.

\begin{theorem}\label{thm-AR-LB}
  Let $G=(V,E,w)$ be a \CWC system with fat links.
  Any algorithm solving \CCR requires
  $\Omega(Z_{\max})$ rounds.
\end{theorem}




\section{Non-Commutative Operators and the Wheel Settings}
\label{sec:wheel}
In this section we consider \CCW for
non-commutative operators in the wheel topology (\figref{fig-ring}).

Trivially, \algref{alg:ccw-hi} and
\theoremref{thm-ccw-covers-ub} 
apply in the non-commutative case if the ordering of the nodes happens
to match an ordering 
induced by the algorithm, but this need not be the case in general.
However, 
it seems reasonable to assume that processing nodes are physically
connected according to their combining order. Neglecting other
possible connections, assuming that the last node is also
connected to the first node for
symmetry, 
and connecting  a cloud node to all processors, we arrive at the \emph{wheel}
topology, which we study in this section. 


Our main result in this section is an algorithm for \CCW for arbitrary
wheel topology that works in time which is a logarithmic factor larger
than optimal.  In contrast to the result of \theoremref{thm-ccw-ub}
that applies only to graphs with fat links, here we analyze the wheel
topology with arbitrary bandwidths (assuming symmetric links).
We note that by using
standard methods \cite{LadnerF-80}, the algorithm presented in this section can be extended to
compute, with the same asymptotic time complexity, all prefix sums,
i.e., compute, for each $0\le j<n$, $\bigotimes_{i=0}^jS_i$.

We distinguish between \emph{holistic} and
\emph{modular} combining operators, defined as follows.
In modular combining, one can apply the combining
operator to aligned, equal-length parts of operands to get the output
corresponding to that part. For example, this is the case with vector
(or matrix) addition: to compute any entry in the sum, all that is
needed is the corresponding entries in the summands.
If the operand is not modular, it is called holistic
(e.g., matrix multiplication).
We show that in the modular case, using pipelining, a logarithmic
factor can be shaved off the running time (more precisely, converted
into an additive term). 


We start by defining the cloud intervals of nodes in the wheel settings.

\begin{definition}\label{def-intervals}
  The \textbf{cloud bandwidth} of a processing node $i \in V_p$ in a
  given wheel graph is $\bc(i)\DEF w(i,v_c)$.  An \textbf{interval}
  $\interval{i}{i\tp k}\DEF\Set{i,i\tp 1,\ldots,i\tp k}\subseteq V$ is
  a path of processing nodes in the ring. Given an interval
  $I=[i,i+k]$, $|I|=k+1$ is its \textbf{size}, and $k$ is its
  \textbf{length}.  The \textbf{cloud bandwidth} of $I$, denoted
  $\bc(I)$, is the sum of the cloud bandwidth of all nodes in $I$:
  $\bc(I)=\sum_{i\in I}\bc(i)$.  The \textbf{bottleneck bandwidth} of
  $I$, denoted $\blmin(I)$, is the smallest bandwidth of a link in the
  interval: $\blmin(I)=\min\Set{w(i,i\tp1)\mid i,i\tp1\in I}$.  If
  $|I|=1$, define $\blmin(I) = \infty$.
\end{definition}

For ease of presentation we consider the ``one sided'' case in which
node $i$ does not use one of its incident ring links.  As we shall
see, this limitation does not increase the time complexity by more
than a constant factor. Hence we consider the case in which node $i$
cannot send messages on its counterclockwise link.



\begin{definition}\label{def-Ii}
	Let $i$ be a processing node in the wheel settings. We define the following quantities for
	clockwise intervals; counterclockwise intervals are defined
	analogously. 
	\begin{compactitem}
		\item \label{def:kc}
		$\kCR{i}$ is the length
		of the smallest interval starting at $i$, for which the product of
		its size by the total bandwidth to the cloud along the interval
		exceeds $s$,  i.e.,
		$$\kCR{i} = \min
          \left(\Set{n}\,\cup\, \Set{k\mid \left(k\tp 1\right) \cdot \bc([i,i\tp k]) \ge
			s} \right)~.
          $$
		\item \label{def:kl}
		$\kLR{i}$ is the length
		of the smallest clockwise interval starting at node $i$, for which the bandwidth
		of the clockwise-boundary link bandwidth is smaller than the total
		cloud bandwidth of the interval, i.e.,
		$$\kLR{i} = \min
          \left(\Set{n}\,\cup\, \Set{k\mid w(i\tp k,i\tp k\tp 1)<\bc(\interval{i}{i\tp k})} \right)~.
          $$
		\item $\kR{i}= \min\Set{\kCR{i}, \kLR{i}}$.
		\item $\IR{i} = \interval{i}{i\tp \kR{i}} $. The interval $\IR{i}$ is
		called the (clockwise) \textbf{cloud interval} of node $i$.
		\item $\ZR{i} = |\IR{i}| + \dfrac{s}{\blmin(\IR{i})} +
		\dfrac{s}{\bc(\IR{i})}$. $\ZR{i}$ is the \textbf{timespan} of the
		(clockwise) cloud interval of $i$.
	\end{compactitem}
\end{definition}

\subsection{The Complexity  of \CW and \CR}

\begin{theorem}\label{theorem:CW_UB_interval}
	Given the cloud interval $I_i$ of node $i$,
	\algref{alg:cw-hi} solves the $s$-bits $\CW_i$ problem in $O(\ZR{i})$ rounds.
\end{theorem}
\Proof
The BFS tree of the interval would be a simple line graph, that is the whole interval.
Note that Step \ref{cw-alg-send-local} requires
$O\left(|\IR{i}| + {\frac{s}{\bLminR{i}}} \right)$ rounds: there are
$s$ bits to send over $\Theta(|I_i|)$ hops with bottleneck bandwidth $\phi(I_i)$.
The rest of the time analysis is the same as in \theoremref{theorem:CW_UB}.
\QED

We have the following immediate consequence.

\begin{theorem}\label{theorem:two_sided_UB}
	Let $Z_i^\ell$ and $Z_i^r$ denote the timespans of the counterclockwise and the clockwise cloud intervals of $i$, respectively. Then $\CW_i$ can be solved in $O(\min(Z_i^\ell,Z_i^r))$ rounds.
\end{theorem}

We now turn to the lower bound.

\begin{theorem}
	\label{theorem:one_sided_LB}
	In the wheel settings, any algorithm for $\CW_i$
	which does not use link $({i\tm 1}, i)$ requires $\Omega(\ZR{i})$ rounds.
\end{theorem}
\Proof
We show that each term
of $\ZR{i}$ is a lower bound on the running time of any algorithm
solving $\CW_i$.
%

%
%
First, note that any algorithm for $\CW_i$
	that does not use edge $({i\tm 1}, i)$ requires 
	$\Omega(\kCR{i}) \ge \Omega \left( \kR{i}\right) = \Omega \left( |\IR{i}|\right)$
	rounds,
due to the exact same arguments as in \lemmaref{lemma:cluster_diameter}.
%
%
%

Next, we claim that	any algorithm for $\CW_i$
	which does not use edge $({i\tm 1}, i)$ requires 
	$\Omega({s}/{\blmin(\IR{i})})$ rounds.
        To see that note first that if $\kR{i}=0$, then
        $\blmin(\IR{i})=\infty$ and the claim is trivial.  Otherwise,
        let $(j,j\tp 1)\in E$ be any link in $\IR{i}$ with
        $\bL{j}{j\tp 1} = \bLminR{i}$.  Note that $j-i<\kR{i}$ because
        $j+1\in I_i$.  Consider the total bandwidth of links emanating
        from the interval $I' \DEF \interval{i}{j} $. Since we assume
        that the link $(i\tm1,i)$ is not used, the number of bits that
        can leave $I'$ in $t$ rounds is at most
        $t\cdot\left(\bc(I')+ \bL{j}{j\tp 1}\right)$.  Notice
        that 
at least $s$ bits have to leave $I'$.
Observe that $\bc(I') \le \bL{j}{j\tp 1}$, because otherwise we would have
$\kLR{i}=j-i$, contradicting the fact that $k(i)>j-i$.
Therefore, any
algorithm $A$ that solves $\CW_i$ in $t_A$ rounds satisfies
$$
s~\le~t_A\cdot\left(\bc(I') + \bL{j}{j\tp 1} \right)
~\le~ 2t_A\cdot \bL{j}{j\tp 1}
~=~ 2t_A\cdot \bLminR{i} ~,
$$
and the claim follows.

Finally, we claim that
any algorithm $A$ for $\CW_i$ which does not use edge $({i\tm 1}, i)$
requires $\Omega \left({s}/{\bC{\IR{i}}} \right)$ rounds.
%
To see that, recall that $\kR{i} = \min(\kCR{i}, \kLR{i})$.  If
$\kR{i}=n$ then $\IR{i}$ contains all processor nodes $V_p$, and the
claim is obvious, as no more than $\bC{V_p}$ bits can be written to
the cloud in a single round.  Otherwise, we consider the two cases: If
$\kR{i} = \kCR{i}$, then $\kR{i} \ge {s}/{\bC{\IR{i}}} - 1$ by
definition, and we are done since $t_A = \Omega(\kR{i})$.
Otherwise, $\kR{i} = \kLR{i}$. Let
us denote $w_R=\bL{i\tp \kR{i}}{i\tp \kR{i}\tp1}$. In this case we
have $w_R < \bc(\IR{i})$.  We count how many bits can leave $\IR{i}$.
In a single round, at most
$\bC{\IR{i}}$ bits can leave through the cloud links, and at most
$w_R$ bits can leave through the local links. Since $A$ solves
$\CW_i$, we must have
$ s ~\le~t_A\cdot\left(\bC{\IR{i}} +w_R\right) ~\le~ 2t_A \cdot
\bC{\IR{i}} ~,$ and hence
$t_A = \Omega\left({s}/{\bC{\IR{i}}} \right)$.  \QED

\begin{theorem}\label{theorem:two_sided_LB}
	Let $Z_i^\ell$ and $Z_i^r$ denote the timespans of the counterclockwise and the clockwise cloud intervals of $i$, respectively. Then $\CW_i$ requires
	$\Omega(\min(Z_i^\ell,Z_i^r))$ rounds in the wheel settings.
\end{theorem}
\Proof
Let $T$ be the minimum time required to perform $\CW_i$.
Due to \lemmaref{lem:schedule-iff-dynamic}, 
we know that there is a dynamic flow mapping with
time horizon $T$ and flow value $s$ from node $i$ to the cloud.
Let $f$ be such a mapping.
We assume that no flow is transferred \emph{to} the source node $i$, as we can modify $f$
so that these flow units would not be sent from $i$ at all until the point
where they were previously sent back to $i$.
Let $s_L$ and $s_R$ be the total amount of flow that is transferred on links
$(i-1,i)$ and $(i,i+1)$, respectively, and
assume \WLOG that $s_R \ge s_L$.
Let $f'$ be a new dynamic flow mapping which is the same as $f$, except that no
flow is transferred on link $(i-1,i)$.
Since $f$ is a valid dynamic flow that transfers all $s$ flow units from $i$ to the cloud,
$f'$ has flow value at least $s-s_L$.
Let $A$ be a schedule derived from $f'$. The runtime of $A$ is at most $T$ rounds.
Let $A'$ be a schedule that runs $A$ twice: $A'$ would transfer $2(s-s_L)$ bits from node $i$
to the cloud. Since $s \ge s_R+s_L$, we get that:
$2(s-s_L) = 2s -2s_L \ge s + s_L + s_R -2s_L \ge s$, and thus $A'$ solves $\CW_i$
without using link $(i-1,i)$.
From \theoremref{theorem:one_sided_LB}, we get a lower bound for $2T$ of
$\Omega(Z_i^r) = \Omega(\min(Z_i^\ell,Z_i^r))$.
\QED

From \theoremref{theorem:two_sided_UB} and \theoremref{theorem:two_sided_LB}
we get the following corollary for the uniform wheel:
\begin{corollary}\label{cor-ub}
	In the uniform wheel topology with cloud bandwidth $\bc$ and local
	link bandwidth $\bl\ge\bc$, \CW can be solved in
	$\Theta\left({s\over\bl}+\min(\sqrt{s\over\bc},{\bl\over\bc})\right)$
	rounds. If $\bl<\bc$, the running time is $\Theta(s/\bc)$ rounds.
\end{corollary}
\Proof
If $\bl<\bc$, $\kLR{i}=0$, $\phi(I_i)=\infty$ and the result follows.
Otherwise, by definition we have $\kCR{i}=\sqrt{s/\bc}-1$ and
$\kLR{i}=\bl/\bc-1$, hence $|\IR{i}|=O(\min(\sqrt{s/\bc},\bl/\bc))$. It
follows that $\bc(\IR{i})=O(\min(\sqrt{s\cdot bc},\bl))$. The result follows by
noting that
$\bLminR{i}=\bl$.
\QED
\noindent
For example, for $\bc=\sqrt{s}$ and $\bl\ge s^{3/4}$, the
running time is $O(s^{1/4})$.

\emph{Remark.}
    Notice that the same upper and lower bounds hold for the $\CR_i$ problem as well.



\subsection{Holistic Combining}
\label{ssec:wheel-holistic}

We are now ready to adapt \theoremref{thm-ccw-covers-ub}
to the wheel settings.

\begin{definition}\label{def:intervals2}
  Given an $n$-node wheel, let $\IR{i}$ be the cloud interval of $i$
  with the smaller timespan (cf.\ \defref{def-Ii}).  Define
  $j_{\max} = \argmax_i\Set{|\IR{i}|}$,
  $j_c = \argmin_i \Set{\bC{\IR{i}}}$, and
  $j_\ell = \argmin_i \Set{\bLminR{i}}$.  Finally, define
  $Z_{\max} = {|\IR{j_{\max}}| + \frac{s}{\bLminR{j_\ell}} +
    \frac{s}{\bC{\IR{j_c}}} }$.
\end{definition}
In words: $j_{\max}$ is the node with the longest cloud interval,
$j_c$ is the node whose cloud interval has the least cloud bandwidth, and
$j_\ell$ is the node whose cloud interval has the narrowest bottleneck.

Let $C$ be the set of all cloud intervals $\IR{i}$.
\begin{theorem}\label{thm-ccw-intervals-ub}
	In the wheel settings,
	\CCW can be solved in
	$O(\Zmax \log n)$ rounds
	by \algref{alg:ccw-hi}
\end{theorem}

The main difference that we can use to our advantage between the wheel case and
the general case, is that in the wheel case,
for any minimal cover $C'$, $\load(C') \le 2$ (see \lemmaref{lem-coverload}).
We can therefore use the cover $C$ to build a minimal cover $C'$,
and then have the multiplexing of \algref{alg:ccw-hi} add only a constant factor
to the runtime.

We do that by adding another preprocessing stage to \algref{alg:ccw-hi},
in which we
select a cover $C' \subseteq C$ such that every
node is a member of either one or two intervals of $C'$.  It is
straightforward to find such a cover, say, by a greedy
algorithm. Concretely, a cover with a minimal \emph{number of intervals} is
found by the algorithm in \cite{min-circle-cover} (in $O(n\log n)$
sequential time). The covers produced by \cite{min-circle-cover}  are sufficient
for that matter, as the following lemma states.
\begin{lemma}
	\label{lem-coverload}
	Let $C$ be a collection of intervals and denote
	$U=\bigcup_{I\in C}I$.  Let $C'\subseteq C$
	be a minimal-cardinality cover of $U$, and let $\load_{C'}(i)=
	|\{I\in C' : I\ni i\}|$. Then for all
	$i\in U$, $1\le \load_{C'}(i)\le 2$.
\end{lemma}
\Proof Clearly $\load_{C'}(i)\ge1$ for all $i\in U$ since $C'$ is
a cover of $U$. For the upper bound, first note that by
the minimality of $|C'|$, there are no intervals $I,I'\in
C'$ such that $I\subseteq I'$,  because in this case $I$ could have
been discarded. This implies that the right-endpoints of intervals in $C'$
are all distinct, as well as the left-endpoints.
Now, assume for 
contradiction, that there exist three intervals
$I,I',I''\in C'$ such that $I\cap I'\cap I''\ne\emptyset$
(i.e., there is at least one node which is a member of all three).
Then $I\cup I'\cup I''$ is a contiguous interval.
Let $l=\min(I\cup I'\cup I'')$ and $r=\max(I\cup I'\cup I'')$.%
\footnote{
	The notation $\max(I)$ for an interval $I$ refers to the
	clockwise-end node of the interval.
}
By counting, one of the three intervals, say  $I$,
has no endpoint in $\Set{l,r}$. But this means
that $I\subseteq I'\cup I''$, i.e., 
we can discard $I$, in contradiction to the minimality of $|C'|$.
\QED

%

As for the ``low levels'' of the algorithm, i.e. Steps \ref{ccw-startloop}--\ref{ccw-endloop},
a little different approach is required instead of running \algref{alg:ccw-low},
due to the limited local bandwidth.
We present the time analysis for these steps with the following lemma:

\begin{lemma}\label{lem-lo}
	Steps \ref{ccw-startloop}--\ref{ccw-endloop}
	of \algref{alg:ccw-hi} terminate in
	$O\left(|I_{j_{\max}}|+\log|I_{j_{\max}}|\cdot{s\over\phi(I_{j_\ell})}\right)$
	rounds in the wheel settings, with $P_I$ stored in the rightmost node of $I$ for each
	interval $I\in C'$. 
\end{lemma}
\Proof
First, we require
that each input $S_i$ is associated with a single
interval in $C'$. To this end, we use the rule that if a node $i$ is a member
in two intervals 
$I$ and $I'$,
then its input $S_i$ is associated with the interval $I$ satisfying
$\max(I)<\max(I')$, and a unit input $\Unit$ is associated with $i$ in
the context of $I'$, where $\Unit$ is  the unit (neutral) operand for
$\otimes$.
Intuitively, this rule means that the overlapping regions in an
interval are associated with the ``left'' (counterclockwise) interval.

Second, we assume that the interval size, and therefore the number
of leaves in its computation tree, is a power of $2$. Otherwise, let
$p=2^{\ceil{\log|I|}}$ (i.e., $p$ is $|I|$ rounded 
up to the next power of $2$). When doing tree computation over an interval
$I$, we extend it (virtually) to a complete binary tree with $p$
leaves, where the leftmost $p-|I|$ leaves have the unit input
$\Unit$. These leaves are emulated by the leftmost node of $I$ (the
emulation is trivial).

Assume now that we wish to compute the combined value of an interval
$I$ whose length is a power of $2$.
Let $S_i^j$ denote the product of $S_i,\ldots,S_j$.
Computation of the combined value
of interval $I$ whose length is a power of $2$ proceeds in {stages},
where each stage $\ell$ computes all level-$\ell$ products in
parallel.  The algorithm maintains the invariant that after $S_i^j$ is
computed, it is stored in node $j$ (by definition, for all
$0\le i< n$, $S_i^i$ is initially stored at node $i$).  The
computation of a stage is performed as follows.

Let $S_i^{j}=S_i^k\otimes S_{k+1}^j$ be a product we wish to compute
at level $\ell$, and let $S_i^k, S_{k+1}^j$ be the values held by its
children. Note that $k+1-i=j-k=2^{\ell-1}$.  The algorithm forwards
$S_{i}^k$ from node ${k}$ to node $j$, which multiplies it by (the
locally stored) $S_{k+1}^j$, thus computing $S_i^j$, which is stored
in node $j$ for the next level. This way, the number of communication
rounds is just the time required to forward $s$ bits from ${k}$ to
$j$. Using pipelining, this is done in
$j-k+{s\over\phi([k,j])}=2^{\ell-1}+{s\over\phi([k,j])} \le
2^{\ell-1}+{s\over\phi(I)}$
rounds,
and hence the total time required to compute the product
of all inputs of any interval $I$ is at most
\begin{equation}
	\label{eq:aw4}
	\sum_{\ell=1}^{\ceil{\log|I|}}\left(2^{\ell-1}+{s\over\phi(I)} \right)
	~\le~ 2\left(|I|+\log|I|\cdot{s\over\phi(I)}\right)~.
\end{equation}

Now back to the proof of \lemmaref{lem-lo}.
Correctness is obvious. As for the complexity analysis:
By \Eqr{eq:aw4},   every interval $B\in C$ completes line
\ref{ccw-local} in
at most
$O\left(|I_{j_{\max}}|+\log|I_{j_{\max}}|\cdot{s\over\phi(I_{j_\ell})}\right)$
rounds.
Multiplying this bound by $\load(C)\le 2$ due to the multiplexing,
we conclude that the loop of
line \ref{ccw-startloop} completes after
$O\left(|I_{j_{\max}}|+\log|I_{j_{\max}}|\cdot{s\over\phi(I_{j_\ell})}\right)$
rounds.
\QED

\begin{lemma}\label{lem-hi}
	In the wheel settings,
	Step \ref{ccw-high} of \algref{alg:ccw-hi}
	terminates in
	$O(\Zmax \cdot \log n)$
	rounds.
\end{lemma}
\Proof
The usage of \algref{alg:ccw-high:node} remains as it were in the fat-links case.
By \lemmaref{ccw-lem-hi}, Step \ref{ccw-high} requires 
$O(\Zmax(C') \cdot \load(C') \cdot \log |C'|)$ rounds.
Noting that $|C'|\le |C| \le n$ and that all intervals in $C'$ are cloud intervals,
we get an upper bound of
$O(\Zmax \cdot 2 \cdot \log n) = O(\Zmax \cdot \log n)$ rounds.
\QED

The proof of \theoremref{thm-ccw-intervals-ub}
is completed as follows.
The correctness of the algorithm is derived from the general case
(\theoremref{thm-ccw-covers-ub}). As for the time analysis:
The ``low levels'' of the algorithm (Steps \ref{ccw-startloop}--\ref{ccw-endloop})
require $O(\Zmax \log n)$ according to \lemmaref{lem-lo},
noting that $|I_{j_{\max}}|\le n$ and 
that  $Z_{\max} = {|\IR{j_{\max}}| + \frac{s}{\bLminR{j_\ell}} +
\frac{s}{\bC{\IR{j_c}}} }$ by definition.
The ``high level'' of the algorithm (Step \ref{ccw-high})
requires $O(\Zmax \cdot \log{n})$ according to \lemmaref{lem-hi}.

All in all, the algorithm terminates in $O(\Zmax \cdot \log{n})$ rounds.
\QED

\para{Remark: Interval containing node $0$.}  In the case that
$\otimes$ is not commutative, the cover $C'$ may need
to be patched: In this case, we require that
the interval that contains node $0$ does not contain node $n-1$.
If this is not the case after computing the cover, let $I \in C'$ be an interval
that contains both. Partition $I$ 
into two subintervals $I=I_L \cup I_R$, where $I_L$ is the part that ends with
node $n-1$, and 
$I_R$ is the part that starts with node $0$. $I_L$ and $I_R$ replace
$I$ in the cover $C'$ for all purposes, except work: any work assigned
to either 
$I_L$ or $I_R$ will be executed by all nodes of $I$. This may incur an
additional constant slowdown due to multiplexing.

We finish with the lower bound.

\begin{theorem}\label{thm-aw-lb}
	Any algorithm for \CCW in the wheel topology requires 
	$\Omega(Z_{\max})$
	rounds. 
\end{theorem}

\Proof Same as in \theoremref{theorem:CCW-LB}, $\Omega(Z_i)$ is a
lower bound for every node $i$, by reduction from $\CW_i$.
Recall that
$\ZR{i} = {|\IR{i}| + {{s}\over{\bLminR{i}}} + {{s}\over{\bC{\IR{i}}}}}$.
For index $j_{\max}$ we get a lower bound of
$\Omega(Z_{j_{\max}}) \in \Omega(|\IR{j_{\max}}|)$.
For index $j_\ell$ we get a lower bound of
$\Omega(Z_{j_N}) \in \Omega({{s}/{\bLminR{j_\ell}}})$.
For index $j_c$ we get a lower bound of
$\Omega(Z_{j_C}) \in \Omega\left( {s}/{\bC{\IR{j_c}} }\right)$.
Summing them all up, gives the desired lower bound. \QED

\emph{Remark.}
    We note that similar upper and lower bounds of $O(\Zmax)$ and $\Omega(\Zmax)$ hold
    for the \CCR problem, similarly to the proofs in \sectionref{sec:CCR}.

\subsection{Modular Combining}
\label{ssec:wheel-modular}

Intuitively, an operator is a modular combining operator if it can
be applied to aligned pieces of the operands and obtain the
corresponding piece of the result. For example, bit-wise operations
are modular, as is vector addition. Formally, we have the
following.
\begin{definition}
	\label{def:modular-op}
	An operator $\otimes:\Set{0,1}^s\times\Set{0,1}^s\to\Set{0,1}^s$  is
	called \textbf{modular} if there exists a partition
	$s=g_1+g_2+\cdots+g_K$, and  $K$ operators
	$\otimes_i:\Set{0,1}^{g_i}\times\Set{0,1}^{g_i}\to\Set{0,1}^{g_i}$
	such that for any $x,y\in\Set{0,1}^s$, $x\otimes y$ is equal to the
	concatenation $(x_1\otimes_1y_1)(x_2\otimes_2 y_2)\cdots 
	(x_K\otimes_K y_K)$ with $x=x_1 x_2 \cdots x_K$, $y=y_1 y_2 \cdots y_K$ and
	$|x_i|=|y_i|=g_i$ for every $i$.
	The \textbf{grain size
		of $\otimes$ in the given partition} is $\max\Set{g_i\mid 1\le i\le
		K}$, and the \textbf{grain size of $\otimes$} is the minimal grain
	size of $\otimes$ over all partitions.
\end{definition}
Note that any $s$-bits binary operator is trivially modular with grain
size at most $s$. However, small grain size facilitates parallelism
and pipelining by applying the operation on small parts of the operands
independently. Note also that the grain size depends on the way we
break down the operands; the best breakdown is the one that minimizes
the grain size.
%
Given such a breakdown,
in case that all (cloud and local) links have bandwidth at least the grain size,
we can use pipelining to
convert the
logarithmic factor of \theoremref{thm-ccw-intervals-ub}
to an additive term, as the following theorem states.
\begin{theorem}\label{thm-aw-modular:appendix}
	Let $G = (V,E)$ be a graph of the wheel topology in the \CWC model.
	Suppose that $\otimes$ is modular with
	grain size $g$,
	and that $w(e) \ge g$ for every link $e\in E$.	
	Then \CCW can be solved in
	$O\left(Z_{\max}+\log n\right)$
	rounds in the wheel settings.
\end{theorem}

Similarly to \corollaryref{cor-ub}, we can get get result for the uniform case.
\begin{corollary} \label{cor:modular-combining}
	In the uniform wheel topology with cloud bandwidth $\bc$ and local
	link bandwidth $\bl$ s.t. $\bl\ge\bc$, \CCW with an operation of
	grain size $g\le\bc$ can be executed in
	$O\left(\frac{s}{\bl}+\min\left(\sqrt{\frac{s}{\bc}},\frac{\bl}{\bc}\right)+\log n\right)$
	rounds.
\end{corollary}

\textbf{Proof of \theoremref{thm-aw-modular}:}
From a high level perspective,
modular combining is the same as holistic
combining, as presented in \theoremref{thm-ccw-intervals-ub} and \algref{alg:ccw-hi}.
All of the preprocessing steps remain unchanged.  The
difference is in the implementation of the low level steps
(\ref{ccw-startloop}--\ref{ccw-endloop}) and
the high level step (\ref{ccw-high}).

For a low level step, we note that the computation tree of \lemmaref{lem-lo}
is no longer
necessary; thanks to the {modularity} of operator $\otimes$, a simple
one-pass procedure suffices.  For interval $I=\interval{\ell}{r}$, the
algorithm forwards $S_\ell$ from node $\ell$ through the local links
all the way to node $r$, and every node along the way combines the
grains it receives with its local corresponding grain,
and forwards the result grain as soon as it is computed.

More formally, let $i\ne \ell$ be some node in $I$.  Let
$x=S_\ell^{i-1}$ be the operand that $i$ should receive from node
$i-1$, and let $y=S_i$. We consider the partition of $x$ and $y$
according to the grains of $\otimes$ as described in
\defref{def:modular-op}. 
For every grain $x_j$ that $i$ fully receives, it
calculates locally the product $x_j \otimes_j y_j$ and proceeds to
send it immediately to node $i+1$.
\begin{lemma}\label{lem:pipe-lo}
	Let $\otimes$ be a modular operator with grain size $g$ as in
	\defref{def:modular-op},
	and assume that $w(e) \ge g$ for every link $e\in E$.
	Using pipelining, line \ref{ccw-local} of
	\algref{alg:ccw-hi} can be computed in an interval $I$ in
	$O\left({\frac{s}{\phi(I)}}+|I|\right)$
	rounds.
\end{lemma}
\Proof
The number of messages required
to send all $s$ bits from a node $j$ to node $j+1$ using only local links is at least
$\ceil{s/\phi(I)}$. An intermediate node can compute the
result of a grain and start forwarding it to the next node after
$O(1)$
steps at most, since a complete grain is needed.
(either a grain was sent in its entirety, or a part of it was
and the rest can be sent in the next round.)
The lemma follows.
\QED

\begin{corollary}\label{cor:pipe-lo}
	In the settings of \lemmaref{lem:pipe-lo},
	line \ref{ccw-local} of
	\algref{alg:ccw-hi} can be computed in
	$O\left({\frac{s}{\phi(I_{j_\ell})}}+|I_{j_{\max}}|\right) \le O(\Zmax)$
	rounds.
\end{corollary}

\bigskip
The pipelining strategy applies also to the tree nodes computed
in line \ref{ccw-high} of \algref{alg:ccw-hi}. Specifically, we need to
change \algref{alg:ccw-high:node} to work grain-by-grain.
Note that we assume the constant-factor multiplexing of
Step \ref{ccw-1} of \algref{alg:ccw-hi}, and thus describe the algorithm
as if all intervals are operating in complete parallelism.
In addition, we assume the bandwidth of all links to be a multiple of $g$
(note that this assumption may only incur a constant factor slowdown,
as a link $e$ passing a message of size $w(e)$ can be simulated
with $2$ rounds of passing messages of sizes
$\floor{w(e)/g}\cdot g$ and $(w(e) \bmod g)$ since $w(e) \ge g$).

Suppose that interval $I$ is
in charge of filling in the value of a tree node $y$ with children
$y_\ell$ and $y_r$.
We assume that the order of reading and writing a string $S$ in \CR
and \CW is sequential, i.e., the first bits of $S$ are retrieved (or
written, resp.) first. We run, in parallel, \CR on $y_\ell$ and $y_r$,
and we also run \CW of $y$ in
parallel to them. We can implement this parallelization using multiplexing of the nodes of $I$,
while maintaining the same round complexity:
the same operations that make up the \CR and \CW actions would still occur,
it is just the order of the schedule that is tweaked.
As soon as a grain of $y_\ell$ and the corresponding grain of
$y_r$ are available, the node of $I$ that read them from the cloud
computes the corresponding grain of the result and writes it using
$\CW$. This way, computing a grain at a tree-node requires
$O(1)$ rounds.
Hence the first grain is computed at the
root after
$O(\log n)$ rounds.
\CW requires $O(\Zmax)$ rounds, 
and hence the last grain is written
$O(Z_{\max})$ rounds later.
Thus we obtain the following result.
\begin{lemma}\label{lem:pipe-hi}
	Let $\otimes$ be a modular operator with grain size $g$ as in
	\defref{def:modular-op},
	and assume that $w(e) \ge g$ for every link $e\in E$.
	Using pipelining, line \ref{ccw-high} of
	\algref{alg:ccw-hi} can be computed in 
	$O\left(Z_{\max}+\log n\right)$
	rounds.
\end{lemma}

\theoremref{thm-aw-modular} follows from \corollaryref{cor:pipe-lo} and
\lemmaref{lem:pipe-hi}.

\begin{theorem}\label{thm-aw-modular}
  Suppose $\otimes$ is modular with grain size $g$, and that
  $w(e) \ge g$ for every link $e\in E$.  Then \CCW can be solved in
  $O\left(Z_{\max}+\log n\right)$ rounds, where
  $\displaystyle Z_{\max} = \max_{i\in V_p}\Set{|\IR{i}|}
  +\frac{s}{\min_{i\in V_p}\Set{\bC{\IR{i}}}} + \frac{s}{\min_{i\in
      V_p}\Set{\bLminR{i}}}~.  $
\end{theorem}

\def\XX{\mathbf{x}}
\def\YY{\mathbf{y}}
\def\ZZ{\mathbf{z}}

\section{\CWC Applications}
\label{sec:applications}


In this \dsection we briefly explore some of the possible applications
of the results shown in this \paper to
two slightly more
involved applications, namely Federated Learning (\subsectionref{sec:federated})
and File Deduplication (\subsectionref{sec:dedup}).


\ignore{
	In this \dsection we briefly explore some of the possible applications
	of the results shown in this \paper.
	Before we start, we note that 
	matrix (or vector) addition over a finite field satisfy the requirements of
	the combining operator analyzed in \sectionref{sec:va} (we need the
	finite field assumption so that the length of the encoding of the
	result of applying the operation
	is the same as the encoding length of each operand). 
	Furthermore, the addition operator is modular, with grain size which is
	the number of bits required to encode an element in the underlying field.
	In the remainder of this section, we describe two slightly more
	involved applications, namely federated learning (\sectionref{sec:federated})
	and file deduplication (\sectionref{sec:dedup}).
}

\subsection{Federated Learning in CWC}
\label{sec:federated}
\ignore{
	
	Here we explain how can the \cwc framework be used in the
	context of federated learning, abbreviated FL henceforth.
}
Federated Learning (FL) \cite{federated-17,federated-cacm} is a distributed Machine Learning
training algorithm,
by which an ML model for some concept is acquired.
The idea is to train over a huge data set that is distributed across many devices such as mobile phones and user PCs, 
without requiring the edge devices to explicitly exchange their data. Thus it gives the end devices some sense of privacy and data protection.
Examples of such data is personal pictures, medical data, hand-writing or speech recognition, etc.

\ignore{
	First, let us briefly review the concept of FL (see
	\cite{federated-17,federated-cacm} for a more detailed description; we
	concentrate on the \emph{horizontal} flavor of FL here). The
	premise is that while labeled data is abundant, much of it is
	private and the owners are not happy to share it (e.g.,
	medical or financial information, or personal photographs). On the other hand,
	effective learning by neural networks 
	requires massive samples of labeled data. To bridge over this gap,
	FL is a framework wherein each user applies the learning process to
	its private data, and the parameters of their
	learned networks are sent to a central trusted server which computes a weighted average 
	of the
	results and sends the combined values back to the users, 
	who proceed to the next iteration (see \cite{federated-17} for more details).
	The goal is to ensure that the only
	information that leaks is the resulting combined network: no
	individually generated information is shared with others.
}

In \cite{federated-secure}, a cryptographic protocol for FL
is presented, under the assumption that any two users can
communicate directly. The protocol of \cite{federated-secure} is
engineered to be
robust against malicious users, and uses cryptographic machinery such as
Diffie-Hellman key agreement and threshold secret sharing.  
We propose a way to do FL using only cloud storage, without requiring an active trusted central server.
Here, we describe
a simple scheme
that is tailored to the fat-links scenario,
assuming that
users are ``honest but curious.'' 

The idea is as follows.
Each of the users has a vector of $m$ weights.  Weights are represented
by non-negative integers from $\Set{0,1,\ldots,M-1}$, so that user
input is simply a vector in $(\mathbb{Z}_M)^m$. Let $\XX_i$ be the vector of user $i$.
The goal of the computation is to compute $\sum_{i=0}^{n-1}\XX_i$ (using addition over $\mathbb Z_M$) and store the result in the cloud.
We assume that $M$ is large enough so that no coordinate in the vector-sum exceeds $M$, i.e., that
$\sum_{i=0}^{n-1}\XX_i=\left(\sum_{i=0}^{n-1}\XX_i\bmod M\right)$.

To compute this sum securely, {we use basic multi-party computation in the \CWC model}.
Specifically, each user $i$
chooses a private random vector $\ZZ_{i,j}\in(\mathbb{Z}_M)^m$ uniformly,
for each of her neighbors $j$,
and sends $\ZZ_{i,j}$ to user $j$.
Then each user $i$ computes
$\YY_i=\XX_i- \sum_{(i,j)\in E} \ZZ_{i,j}+ \sum_{(j,i)\in E} \ZZ_{j,i}$,
where addition is modulo $M$. 
Clearly, $\YY_i$ is uniformly distributed even if $\XX_i$ is known.
Also note that $\sum_i\YY_i=\sum_i\XX_i$. Therefore all that remains
to do is to compute $\sum_i\YY_i$, which can be done by invoking \CCW,
where the combining operator is vector addition over $(\mathbb{Z}_M)^m$.
We obtain the following theorem from \theoremref{thm-ccw-ub}.

\begin{theorem}\label{thm-FL}
	In a fat-links network,
	an FL
	iteration with vectors in $(\mathbb{Z}_M)^m$ can be computed in
	$O(\Zmax \log^2 n)$ rounds.
\end{theorem}

Since the grain size of this operation is $O(\log M)$ bits, 
we can apply the pipelined version (\corollaryref{cor:modular-combining}) of \CCW
in case that the underlying topology is a cycle,
to obtain the following. 
\begin{theorem}\label{thm-FL=wheel}
	In the uniform $n$-node wheel,  an FL
	iteration with vectors in $(\mathbb{Z}_M)^m$ can be computed in
	$O(\sqrt{(m\log M)/\bc}+\log n)$ rounds, assuming that
	$\bc m\log M\le\bl^2$ and $\bc\ge\log M$.
\end{theorem}
\Proof Using the notation of \sectionref{sec:va} and \subsectionref{sec:wheel}, the assumption implies that $s=m\log M$, and $Z_{\max}=O(\sqrt{s/\bc})=O(\sqrt{(m\log M)/\bc})$. The result follows
from \theoremref{thm-aw-modular} and \corollaryref{cor:modular-combining}.
\QED



\def\dcup{\widetilde{\cup}}

\subsection{File Deduplication With the Cloud}
\label{sec:dedup}

Deduplication, or Single-Instance-Storage (SIS), is a central problem
for storage systems (see, e.g.,
\cite{dedup,dedup-performance,bolosky2000single}). 
Grossly simplifying, the motivation is the following: Many of the
files (or file parts) in a storage system may be unknowingly
replicated. The general goal of deduplication (usually dubbed dedup)
is to identify such replications and possibly discard
redundant copies.  Many cloud storage systems use a dedup mechanism
internally to save space. Here we show how the processing nodes can cooperate to
carry out dedup without active help from the cloud, when the files
are stored locally at the nodes (cf.\
serverless SIS~\cite{serverless-SIS}). We ignore privacy
and security concerns here.




We  consider the following setting. 
Each node $i$ has a set of local files
$F_i$ with their hash values, and the goal is to identify, for each
unique file $f\in\bigcup_i F_i$, a single owner user $u(f)$.
(Once the operation is done,  users may delete any file they do not
own.)

This is easily done with the help of \CCW as follows.
Let
$h$ 
be a hash function. 
For file $f$ and processing node $i$, call the pair $(h(f),i)$ a
\emph{tagged hash}. The set
$S_i=\Set{(h(f), i) \mid f\in F_i}$ of tagged hashes of $F_i$ is the input of  node $i$.
Define the operator $\dcup$ that takes two sets $S_i$
and $S_j$ of tagged hashes, and returns a set of tagged hashes without duplicate hash values, i.e., if $(x,i)$ and $(x,j)$ are both in the union $S_i\cup S_j$,
then only $(x, \min(i,j))$ will be in $S_i\,\dcup\,S_j$.
Clearly
$\dcup$ is associative and commutative, has a unit element ($\emptyset$), and
therefore can be used in the \CCW algorithm.
Note that if  the total number of unique files in the system is $m$, then
$s=m\cdot (H+\log{n})$.
Applying \CCW with operation $\dcup$ to inputs $S_i$, 
we obtain  a set of tagged hashes $S$ for all files in the system,
where $(h(f),i) \in S$ means that user $i$ is the owner of file $f$.
Then we invoke \CCR 
to disseminate the ownership information to all nodes.
Thus dedup can be done in \cwc in $O(\Zmax \log^2 n)$ rounds.

\ignore{
	\subsection{Cloud File Deduplication}
	Next, assume that a \emph{multiset} $F$ of files is
	stored in the cloud, 
	and
	the goal is to replace each duplicate file by
	a reference to  its remaining copy.
	To do so, we partition the files among the users so that  each node
	$i$ is assigned a subset, $F_i\subseteq F$ whose size $|F_i$ is
	proportionate to 
	$\bc(i)$ (we assume that for all $i$,  $\bc(i)\ge H$).
	\boaz{how is this done?!}
	Then  each node $i$ reads $h(f)$ for all $f\in F_i$ from the cloud,
	and proceed to the algorithm described in \ref{local-file-dedup}
	with the following difference: instead of tagging the hash of a file
	with a node identifier, hashes are tagged by a reference to its storage
	in the cloud.
	After computing the result $S$,  each user $i$
	deletes from the cloud the files of $F_i$  files whose hashes in $S$
	are  not tagged by their location in $F_i$, and we are done.
}


\section{Conclusion and Open Problems}
\label{sec:conc}

In this paper we have introduced  a new model that incorporates cloud
storage with a bandwidth-constrained communication network. We have developed
a few building blocks in this model, and
used these primitives to obtain effective  solutions to some
real-life distributed applications.
There are many possible directions for future work; below, we mention a few.
%

One interesting direction is to validate the model with
\emph{simulations and/or implementations} of the algorithms, e.g., implementing the
federated learning algorithm suggested here.

A few algorithmic question are left open by this paper.
For example, can we get good approximation ratio for the problem of combining in a general (directed, capacitated) network? Our results apply to fat links and the wheel topologies.

Another interesting issue 
is the case of \emph{multiple cloud
nodes}: How can nodes use them effectively, e.g., in combining?
Possibly in this case one
should also be concerned with privacy considerations.


Finally, 
\emph{fault
  tolerance}: Practically, clouds are considered highly reliable.
How should we exploit this fact to build more robust systems? and on the other
hand, how can we build systems that can cope with varying cloud
latency?



\newpage

\bibliographystyle{plain}
\bibliography{cwc}

\end{document}